\UseRawInputEncoding
\documentclass[twocolumn, twocolappendix]{aastex631}

\definecolor{MyDarkBlue}{rgb}{0,0.08,0.5}
\definecolor{MyDarkRed}{rgb}{0.7,0.02,0.02}
\definecolor{MyDarkmagenta}{rgb}{0.0,0.7,0.0}

\usepackage{comment}
\usepackage{tabularx}
\usepackage{booktabs}
\usepackage{xcolor}
\usepackage{amsmath}

\accepted{August 19, 2025}

\submitjournal{ApJ}
\shortauthors{Yamaguchi et al.}

\newcommand{\asiaa}{Academia Sinica Institute of Astronomy and Astrophysics, 11F of ASMA Building, No.1, Sec. 4, Roosevelt Rd, Taipei 106216, Taiwan}

\newcommand{\kyushu}{Department of Earth and Planetary Sciences, Faculty of Sciences, Kyushu University, Nishi-ku, Fukuoka 819-0395, Japan}

\newcommand{\naoj}{National Astronomical Observatory of Japan, National Institutes of Natural Sciences, 2-21-1 Osawa, Mitaka, Tokyo 181-8588, Japan}

\newcommand{\nsyu}{Department of Physics, National Sun Yat-Sen University, No. 70, Lien-Hai Road, Kaohsiung City 80424, Taiwan, R.O.C.}
\newcommand{\cag}{Center of Astronomy and Gravitation, National Taiwan Normal University, Taipei 116, Taiwan}

\begin{document}

\title{Peculiar Disk Substructures Associated with the Young Eruptive Star EX Lupi}

\correspondingauthor{Masayuki Yamaguchi}
\email{yamaguchi.masayuki.376@m.kyushu-u.ac.jp}
\author[0000-0002-8185-9882]{Masayuki Yamaguchi}
\affil{\asiaa}
\affil{\kyushu}
\affil{\naoj}
\author[0000-0003-2300-2626]{Hauyu Baobab Liu}
\affil{\nsyu}
\affil{\cag}
\author[0000-0001-9248-7546]{Michihiro Takami}
\affil{\asiaa}
\author[0000-0002-5067-4017]{Pin-Gao Gu}
\affil{\asiaa}

\begin{abstract}
Young eruptive stars such as EXors undergo dramatic accretion outbursts characterized by sudden optical brightenings, yet the underlying physical mechanism remains uncertain. We present high-resolution Atacama Large Millimeter/submillimeter Array (ALMA) Band 3 and 4 continuum observations of EX Lupi, the prototypical EXor-type variable, reconstructed using super-resolution imaging with sparse modeling. Our images reveal, for the first time, two distinct substructures: a compact, crescent-shaped inner arc within 10 au of the star, and a narrow outer ring at 30 au. The inner arc is strongly elongated and casts a shadow observed in the Very Large Telescope/SPHERE near-infrared scattered light. The outer ring exhibits a radial width comparable to the local pressure scale height, consistent with moderately efficient dust trapping. Geometric and thermal analysis of the disk surface, based on combined ALMA and SPHERE data, indicates that the disk is moderately flared with an average disk temperature consistent with that of classical T Tauri disks. The observed substructures suggest dynamical perturbations-plausibly induced by a massive companion companion-that may modulate accretion rates through gravitational interaction with the inner arc. These findings provide morphological evidence linking disk substructure to episodic accretion in the structurally mature disk.

\end{abstract}



\section{Introduction} \label{sec:intro}
In the past decade, high-resolution observations using the Atacama Large Millimeter/submillimeter Array (ALMA) have revealed that many protoplanetary disks around young stars exhibit substructures$-$rings, gaps, spirals, and arcs$-$that deviate from smooth disk models \citep[e.g.,][]{yen_gas_2016, tang_planet_2017, kudo_spatially_2018, tsukagoshi_discovery_2019, hashimoto_asymmetric_2021, liu_first_2024,yamaguchi_alma_2024, shoshi_alma_2025}. These substructures may influence mass transport within the disk and onto the central star, potentially linking disk morphology to accretion processes. A particularly striking manifestation of variable accretion is found in EXor-type stars, a subclass of eruptive young stars that experience episodic accretion outbursts \citep{audard2014}. These events are marked by optical brightenings of $1-6$ mag and a temporary increase in mass-accretion rates from $\sim10^{-9}$ to $10^{-7}~M_{\odot}\mathrm{~yr}^{-1}$ \citep{Fischer2023ASPC}. The physical origin of these bursts remains uncertain, with proposed mechanisms including thermal and gravitational instabilities \citep{zhu_two-dimensional_2009, machida_cloud_2024}, disk fragmentation \citep{vorobyov_variable_2015}, and dynamical perturbations induced by an embedded companion \citep{nayakshin_youngest_2024}. Identifying morphological signatures associated with such mechanisms is therefore crucial for understanding how disk structure regulates episodic accretion.

EX Lupi, the prototypical EXor-type variable, serves as an ideal laboratory for studying episodic accretion phenomena. This M0.5-type star, located at a distance of $154.7 \pm 0.4$ pc \citep{Gaia2023}, has a quiescent stellar luminosity of $0.43~L_{\odot}$ \citep{Wang2023ApJ} and a dynamical stellar mass of $0.5~M_{\odot}$ \citep{hales_circumstellar_2018}. It has undergone three major accretion outbursts$-$in 1945, 1955, and 2008$-$each marked by an optical brightening of $\sim5$ mag \citep{mclaughlin_nova-like_1946, Herbig1977, aspin_2008_2010}. In addition to these large-scale events, EX Lupi exhibits recurrent smaller outbursts during its quiescent phases \citep{herbig_ex_2007, de_miera_brightness_2023}, indicating a persistent state of variability. These recurring episodes have prompted extensive multiwavelength monitoring campaigns aimed at identifying the physical mechanisms responsible for the observed accretion dynamics.

Recent observations have begun to unveil the dynamical complexity of an inner disk region of EX Lupi. Near-infrared spectroscopy has revealed rapid changes in the accretion funnel geometry during outburst phases, implying structural evolution on sub-au scales \citep{singh_accretion_2024}. Magnetic field measurements show multikilogauss field strengths capable of sustaining magnetospheric accretion \citep{pouilly_multi-kilogauss_2024}. In quiescence, mid-infrared spectroscopy indicates vertical chemical stratification and thermal restructuring in the disk atmosphere \citep{kospal2023, woitke_2d_2024}. These results consistently suggest that the inner disk region is dynamically active and potentially central to triggering episodic mass accretion.

Previous ALMA observations have identified a compact and radially concentrated dust disk \citep{hales_circumstellar_2018, white_alma_2020}. However, those studies lacked the spatial resolution needed to directly image small-scale disk substructures that could link morphology to accretion behavior. To address this gap, we present high-resolution ALMA Band 3 and 4 continuum observations of EX Lupi, reconstructed using the sparse-modeling-based super-resolution technique $\tt PRIISM$ \citep{Nakazato2020, Nakazato_priism_2020}. We present the first spatially resolved detection of a compact, crescent-shaped inner arc and an outer ring.

The rest of the paper is organized as follows. In Section~\ref{sec:observation}, we describe the ALMA observations and outline the PRIISM super-resolution imaging procedure. In Section~\ref{sec:disk_str_prop}, we characterize the morphology of the outer ring and inner arc structures. Section~\ref{sec:nearinfraded_image} analyzes the disk-surface geometry using scattered-light data from Very Large Telescope (VLT)/SPHERE. Section~\ref{sec:disk} derives the thermal structure and radial extent of the dust disk. In Section~\ref{sec:discussion}, we discuss the implications of our findings for disk evolution and the physical origins of episodic accretion. Appendices~\ref{appendix:clean_images} and \ref{sec:priism_image} present CLEAN-based imaging comparisons and performance assessments of the PRIISM method. Appendix~\ref{appendix:ellipse_fit} describes the ellipse fitting procedure used to estimate the disk-surface geometry.

\section{Observations and Imaging Procedures}\label{sec:observation}

In this section, we describe the ALMA observations and subsequent imaging procedures for EX Lupi. Section~\ref{sec:obs_datareduction} outlines the observational setup and manual data reduction process. Section~\ref{sec:clean_imaging} details the self-calibration and standard CLEAN imaging technique used to enhance image fidelity. In Section~\ref{sec:imaging_priism}, we introduce the PRIISM imaging technique, a super-resolution imaging method based on sparse modeling, which provides improved spatial resolution and reliable disk morphology compared to CLEAN.

\subsection{Data Reduction}\label{sec:obs_datareduction}

EX Lupi was observed with ALMA Bands 3 and 4 (project code: 2017.1.00388.S, PI: Hauyu Baobab Liu). The observations were performed in 2017 November 6 (Band 4) and 2017 November 7 (Band 3) during the quiescent phase before a large outburst in 2022 February \citep{Wang2023ApJ}. The observations were performed with a 12 m array consisting of a C40-9 antenna configuration with the baseline length extending from 113 to 13,894 m. The correlators are configured for continuum observations to cover four 1.88 GHz-wide spectral windows. We used the Common Astronomy Software Applications package \citep[$\tt CASA$, version 5.1.1;][]{CASATeam2022} for the calibration. The data were initially calibrated with the $\tt CASA$ pipeline reduction scripts provided by the ALMA Regional Centers. For the Band 3 dataset, the absolute flux scale and bandpass shape were corrected with J1427-4206, and the phase variations were corrected with J1610-3958. For the Band 4 dataset, the absolute flux and bandpass calibration were applied with J1617-5848, and the phase calibration was applied with J1610-3958. Each final continuum dataset was produced by combining the four continuum spectral windows, resulting in an average frequency of 92.95 GHz (3.2 mm) for Band 3 and 152.95 GHz (2.0 mm) for Band 4.

We determined the common phase center as the local peak of the inner arc emission around the center of the $\tt tclean$ image  (hereafter CLEAN image) using the Band 4 dataset. After shifting the phase center of each band's data set using the $\tt CASA$ task $\tt fixvis$, all datasets were aligned to the common phase center using $\tt fixplanets$, adopting the International Celestial Reference System (ICRS) coordinate of (16$^{\mathrm{h}}$03$^{\mathrm{m}}$5$^{\mathrm{s}}$.476, $-40^{\circ}18'25\farcs83$). We then calculated the maximum recoverable scale (MRS) given by $0.983~\lambda/D_{5} $ (in radians) for each observed data set, where $\lambda$ is the observing wavelength and $D_{5}$ is the 5th percentile of the $uv$ distance (Equation (7.7) in ALMA Technical Handbook). The MRSs for Bands 3 and 4 provided $1\farcs18$ and $0\farcs66$, respectively.

\subsection{Self-calibration and CLEAN Imaging}\label{sec:clean_imaging}
To improve the signal-to-noise ratio (SNR) of the images by correcting a systematic gain error, we apply a phase-only self-calibration ($\tt calmode = p$) with a CLEAN model as a reference calibrator for each dataset. The CLEAN model was produced by using multi-frequency synthesis \citep[$\tt nterm = 1$;][]{rauMultiscaleMultifrequencyDeconvolution2011} and multi-scale approach \citep{cornwell_multiscale_2008} with scale parameters of $[0, 1, 3] \times \theta_{\rm cl}$ (where $\theta_{\rm cl}$ is the CLEAN beam size) on $\tt CASA$ version 6.5.0. In the self-calibration process, we employed a Briggs robustness parameter of 0.5 for the imaging and CASA task $\tt gaincal$ to average whole scans ($\tt solint = inf$). These processes for the Band 3 and 4 datasets provided SNRs in the final reconstructed images of 8 and 11, respectively, which reached slight improvements of 1.2 in SNR compared with a non-self-calibrated CLEAN image and mitigated patchy artifacts outside the region of source emission. The final restored CLEAN images provided a beam size of $0\farcs103 \times 0\farcs068 ~ (\rm PA = -77^{\circ})$ for Band 3 and $0\farcs064 \times 0\farcs042 ~ (\rm PA = 84^{\circ})$ for Band 4. The rms noise values, which are measured in off-source areas, for Bands 3 and 4 were $36~\mu \rm Jy~beam^{-1}$ and $46~\mu \rm Jy~beam^{-1}$, respectively. The obtained sensitivity levels were satisfactory, with ideal point-source sensitivities of $\sim 30-40~\mu \rm Jy~beam^{-1}$ given by the observing setup (Equation (9.8) in the ALMA Technical Handbook). Further description of the CLEAN images is given in Appendix \ref{appendix:clean_images}, and the corresponding images are shown in Figure \ref{fig:clean_images}.

\subsection{PRIISM Imaging}\label{sec:imaging_priism}

As an alternative to standard CLEAN imaging, we employ $\tt PRIISM$\footnote[1]{Python Module for Radio Interferometry Imaging with Sparse Modeling ($\tt PRIISM$) is a public ALMA imaging tool based on sparse modeling, available at \url{https://github.com/tnakazato/priism}} \citep[version 0.11.5;][]{Nakazato2020, Nakazato_priism_2020}. This method produces smoother model images and achieves improved effective spatial resolution compared to CLEAN, owing to its regularization scheme that better preserves intrinsic disk morphology and suppresses spurious patchy features. More specifically, this approach involves regularized maximum-likelihood optimizations with $\ell _1$+TSV imaging and a cross-validation (CV) scheme as illustrated in \citet{yamaguchi_super-resolution_2020}. The model image is reconstructed by minimizing a cost function consisting of the chi-squared error between the visibility model derived from the model image and the observed visibility, accompanied by two regularization terms, namely $\ell_1$ norm and the total squared variation (TSV) controlled by the hyper-parameters $\Lambda_{l}$ and $\Lambda_{tsv}$, respectively. 

With the self-calibrated visibility data in Section \ref{sec:clean_imaging}, we obtain one model image that minimizes the cost function for a given set of $(\Lambda_{l}, \Lambda_{tsv})$ and have a set of model images with a range of values of $(\Lambda_{l}, \Lambda_{tsv})$ \footnote[2]{We set up parameter ranges for $(\Lambda_{l}, \Lambda_{tsv}$) of [($8\times10^{4}, 9\times10^{4}, 1\times10^{5}, 2\times10^{5}, 3\times10^{5}$), ($1\times10^{12}, 2\times10^{12}, 3\times10^{12}, 4\times10^{12}, 5\times10^{12}$)] for Band 3 and [($1\times10^{4}, 3\times10^{4}, 5\times10^{4}, 7\times10^{4}, 9\times10^{4}$), ($1\times10^{12}, 3\times10^{12}, 5\times10^{12}, 7\times10^{12}, 9\times10^{12})$] for Band 4.}. To select a pair of $(\Lambda_{l}, \Lambda_{tsv})$ for the optimal image, we employ a 10-fold CV approach. The model with minimum cross-validation error (CVE) is set to the optimal one \citep{yamaguchi_super-resolution_2020}. The final pairs of $(\Lambda_{l}, \Lambda_{tsv})$ with the minimum CVE provided ($10^{5}$, $3\times10^{12}$) for Band 3 and ($7\times10^{4}$, $3\times10^{12}$) for Band 4, respectively. The final model image has units of janskys per pixel ($\rm Jy~pixel^{-1}$) because this model imaging process does not include beam convolution.

In Appendix~\ref{sec:priism_image}, we present a quantitative evaluation of PRIISM imaging performance, including (i) validation of image fidelity in the visibility domain, (ii) estimation of effective spatial resolution using a point-source injection method, and (iii) a procedure for generating restored images that incorporates beam convolution and residual addition for compatibility with conventional imaging outputs.

Throughout this study, we adopt a dual-image strategy that leverages the strengths of both high-resolution modeling and reliable noise estimation. The original PRIISM model images are used for morphological analyses where the preservation of spatial resolution is critical. In contrast, the restored images serve as beam-matched representations suited for analyses requiring accurate noise characterization, such as spectral index measurements.

\section{Peculiar Disk Structures and Their Properties}\label{sec:disk_str_prop}

\begin{figure*}[t]
\centering
\includegraphics[width=0.98 \textwidth]{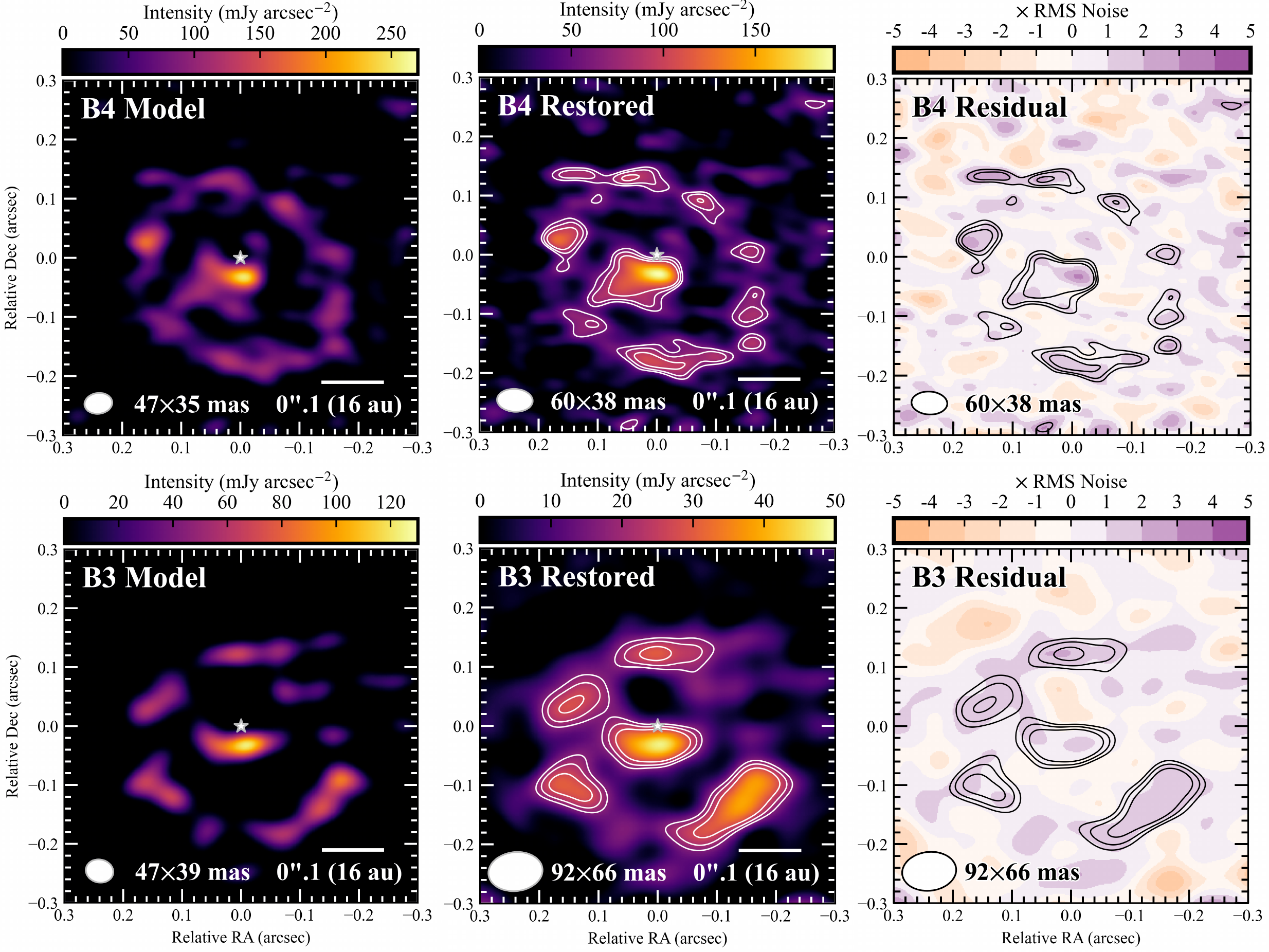}
\caption{Gallery of ALMA (PRIISM) continuum images and their residual maps at 3.2 mm / 93 GHz (Band 3: bottom row) and 2.0 mm / 153 GHz (Band 4: top row) of the EX Lupi disk. The same linear color scale is adopted. A white bar of $0\farcs1$ (= 16 au) is provided to reference the angular scales. The central stellar symbol ($\star$) indicates the stellar position measured in Gaia observations. Left: Model image reconstructed using PRIISM imaging. The image initially has a unit of janskys per pixel ($\rm Jy~pixel^{-1}$; Section \ref{sec:imaging_priism}), while it is converted to millijanskys per square arcsecond ($\rm mJy~arcsec^{-2}$) for comparison with the restored image (middle). The filled white ellipse denotes the effective spatial resolution $\theta_{\rm eff}$. The resolution is estimated from an artificial point-source injection method (Appendix \ref{sec:effective_resolution}). Middle: restored image after convolution with an elliptical Gaussian representing the main lobe of the synthesized beam, along with adding the dirty map of residuals (right). Each white contour corresponds to $[3, 4, 5]\times \sigma_{\rm noise}$, where $\sigma_{\rm noise}$ is the RMS noise value calculated from the emission-free area ($36~\mu \rm Jy~\rm beam^{-1}$ for Band 3 and $48~\mu \rm Jy~\rm beam^{-1}$ for Band 4). The restored image is expressed in a unit of $\rm Jy~beam^{-1}$, while being converted to $\rm mJy~arcsec^{-2}$. Right: dirty map of the residuals, obtained by subtracting the model from the observation in the visibility domain. The unit is expressed in image RMS noise. The black contours represent the same as the white ones on the restored image (middle).}
\label{fig:priism_images}
\end{figure*}

Figure \ref{fig:priism_images} shows the Band 3 and 4 images obtained using the PRIISM imaging and analysis. The left panels show the model image without beam convolution of the observations (Section \ref{sec:imaging_priism}), the middle panels show the restored images that include the effect of beam convolution of the observations (Appendix \ref{sec:restored_image}), and the right panels show the residuals between the two images (Appendix \ref{sec:restored_image}).

The model and restored images, which represent the spatial distributions of the thermal dust continuum, exhibit the following features in both Band 3 and 4 images.

Emission clumps form a ring (hereafter ``outer ring'') elongated from the northeast to southwest on a 0\farcs4 scale. Despite the modest SNR for these clumps (3$\sigma$-5$\sigma$ levels), the location of the ring in the Band 3 and 4 images appears to approximately match, confirming its presence. Despite differences in angular resolution and SNR, the locations of the clumps along the east-to-north directions appear consistent between the Band 3 and 4 images. In contrast, the Band 3 image shows a bright clump in the southwest which is not seen in the Band 4 image. This discrepancy between the two bands may be due to frequency variations in dust opacity, but observations with higher SNR are required to confirm or reject the presence of this clump. The integrated fluxes of the ring, measured from regions with emission exceeding the $3\sigma$ threshold in the restored images, are $1.0\pm0.2$ mJy in Band 3 and $2.1\pm0.4$ mJy in Band 4.

Near the stellar position, $r\sim$0\farcs02 from the star, we find a compact emission component elongated in the east-west direction on a $\sim$0\farcs1 scale, surrounding the star like an arc (hereafter the ``inner arc''). In the Band 4 image only, the inner arc emission is associated with a tail-like feature extending toward the southeast to $\sim$10 au from the star. Again, the absence of this tail in the Band 3 image may be due either to its low SNR or to frequency variations in the dust opacity.

The rest of the section is organized as follows. In Section \ref{sec:outer_ring}, we analyze the geometrical and physical properties of the outer ring using the restored Band 4 image, which provides a clearer ring structure with higher spatial resolution than the Band 3 image. In Section \ref{sec:inner_disk}, we describe the detailed properties of the inner arc. 

\subsection{Outer Ring}\label{sec:outer_ring}

The outer dust ring identified in the ALMA continuum images represents a prominent substructure in the EX Lupi disk, providing a valuable diagnostic of the physical conditions and dynamical processes operating in the outer disk regions. To characterize this feature in detail, we divide our analysis into two parts. In Section~\ref{sec:outerring_geometry}, we first investigate the geometric properties of the ring, including its inclination, position angle (PA), and radial width, based on ellipse fitting and radial profile analysis. In Section~\ref{sec:dust_trapping}, we assess whether the observed narrow width of the ring is consistent with dust trapping in a pressure maximum by comparing its radial confinement to the local pressure scale height. Together, these analyses allow us to evaluate the physical origin of the outer ring structure.

\subsubsection{Detailed Geometry}\label{sec:outerring_geometry}

\begin{figure*}[t]
\centering
\includegraphics[width=0.95 \textwidth]{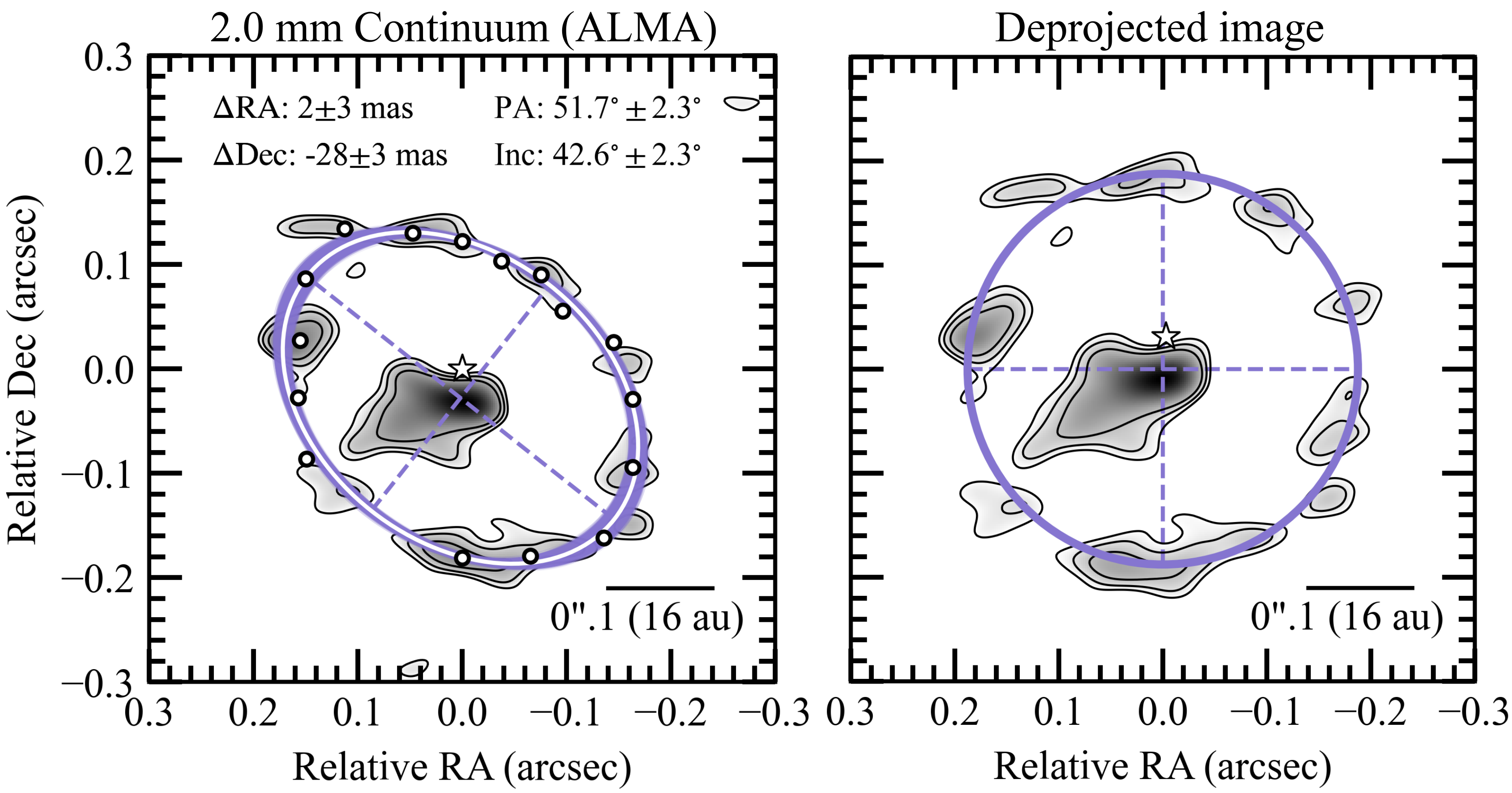}
\caption{Left: the best-fit ellipse to the outer ring overlaid on the restored Band 4 image, with contour levels at $[3, 4, 5] \times \sigma^{\rm B4}_{\rm noise}$, where $\sigma^{\rm B4}_{\rm noise}$ is the RMS noise of $48~\mu \rm Jy~ beam^{-1}$. The purple curves represent the distribution of solutions considering the estimated $1\sigma$ error. The white dots along the outer ring indicate the radially averaged peak positions, spaced at $20^{\circ}$ intervals. The outer peak in the PA of $140^{\circ} \sim 160^{\circ}$ could not be identified due to insufficient angular resolution or low SNR and was therefore excluded from the fit. Right: the deprojected image with the inclination ($=42^{\circ}.6$) and PA ($=51^{\circ}.7$) derived from the best-fit ellipse. The center of the ellipse fit is used as the center position for deprojection. The purple circle indicates the best-fit model, while the white star symbol indicates the position of EX Lupi.}
\label{fig:ellipse_fit}
\end{figure*}


\begin{figure}[hbtp!]
\centering
\includegraphics[width=0.4 \textwidth]{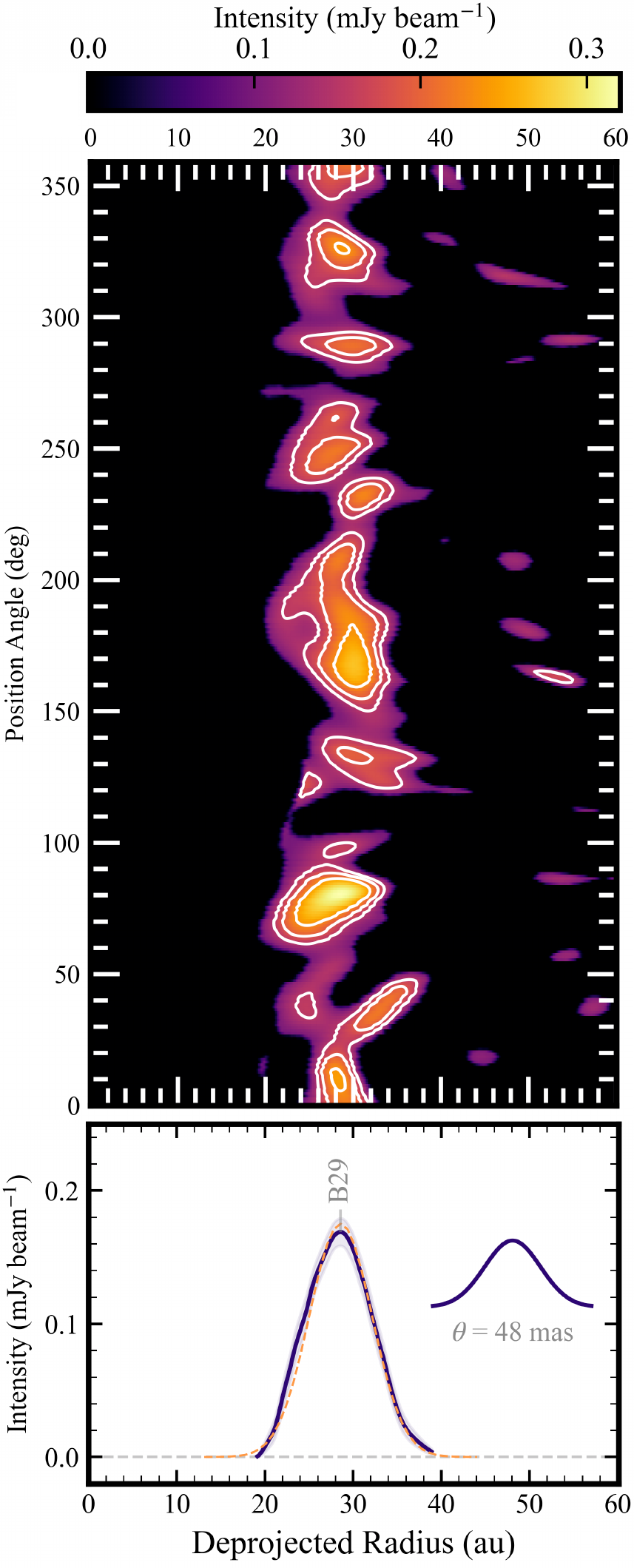}
\caption{Top panel: deprojected PA profile of the outer ring of EX Lupi, derived from the Band 4 restored image. The contour levels are $[3, 4, 5] \times \sigma_{\rm noise}$. Note that the emission from the inner arc ($r < 20$ au) is removed. The center coordinates are set based on an ellipse fit to the outer ring. Bottom panel: Radial intensity profile averaged over the full azimuthal angle, displayed on a linear color scale. The profile (solid purple curve) is linearly interpolated onto radial grid points spaced by 0.1 au using $\tt interpolate.interp1d$ from the $\tt SciPy$ module. The light purple ribbon represents the error of the mean at each radius. The dashed orange curve corresponds to the best-fit Gaussian profile to the observed data. The geometric mean of the spatial resolution is shown in the inset at the top-right corner.}
\label{fig:radialprofile}
\end{figure}

Assuming that the deprojected outer ring is circular, we first derive its PA and inclination by performing an ellipse fit in the same manner as in \cite{yamaguchi_alma_2021}. The left panel of Figure \ref{fig:ellipse_fit} shows the best-fit ellipse overlaid on the Band 4 restored image, yielding a PA of $51^{\circ}.7 \pm 2^{\circ}.2$ and an inclination of $42^{\circ}.6 \pm 2^{\circ}.4$. This inclination value is in reasonable agreement with the $38^{\circ} \pm 4^{\circ}$ derived from a Keplerian model rotation of the ALMA $^{12}\text{CO}~(J= 2-1)$ gas kinematics \citep{hales_circumstellar_2018}. The center of the ellipse has an offset of $\sim 5$ au or $(\Delta \text{R.A.}, \Delta \text{decl.}) = (31,~28)~\text{mas}$ from the central stellar position determined using Gaia DR 3 parallax measurements \citep{Gaia2023} \footnote[3]{The stellar position is determined using Gaia DR3 parallax \citep{Gaia2023}, incorporating its proper motion correction of $(-23$ mas yr$^{-1}$ in Right Ascension, $-11$ mas yr$^{-1}$ in Declination) to align with the ALMA observing epoch. The resulting ICRS coordinates for EX Lupi are ($16^{\rm h}03^{\rm m}5^{\rm s}.476$, $-40^{\circ} 18' 25\farcs764$).}. The interpretation of this offset is discussed in Section \ref{sec:origin_innerarc}.

Next, we investigate the radial distribution of the outer ring. Figure \ref{fig:radialprofile} shows the azimuthally averaged radial intensity profile of the outer ring, $I_{\nu}(r)$, which was obtained by deprojecting the Band 4 restored image using  inclination and PA geometric parameters. The origin at $r=0$ au is based on the center of the elliptical model, which is offset by 5 au from the stellar position. We then exclude the inner arc region (where $r < 15$ au) on the profile. The uncertainty of the intensity $\hat{\sigma}_{I}$ is measured for each radius by calculating the standard error over the entire PA range, treating the beam size $\theta$ as the smallest independent unit. This is expressed as $\hat{\sigma}_{I}=\sigma_{I}/\sqrt{N_{\rm B}}$, where $\sigma_{I}$ is the standard deviation of the brightness within the concentric ring at $r = r_{i}$, and $N_{\rm B} = 2 \pi r_{i} /\left\langle\theta\right\rangle$ represents the number of beams within the ring. Here, $\left\langle\theta\right\rangle$ ($= 0\farcs05$) is defined as the geometric mean of the beam size.

The outer ring is found to be only marginally resolved in the radial direction relative to the beam size. To quantify its intrinsic width, we first fit the radial intensity profile with the following Gaussian function:

\begin{equation}
I_{\nu}^{\text{model}}(r) = I_{\rm ring} \exp \left\{ -\frac{\left(r - r_{\rm ring}\right)^{2}}{2 w_{\rm d}^{2}} \right\},
\end{equation}
\noindent
where $I_{\rm ring}$ and $r_{\rm ring}$ represent the peak intensity of the outer ring and its radius from the center, and $w_{\rm d}$ denotes the ring width. To determine the best-fit values for the free parameters ($I_{\rm ring}$, $r_{\rm ring}$, and $w_{\rm d}$), we employed a Markov Chain Monte Carlo (MCMC) fitting procedure using the \texttt{emcee} sampler \citep{Foreman2013}. We adopted uniform (flat) priors and ran 100 walkers for 500 steps, discarding the initial 250 steps as burn-in. The resulting posterior distributions provided median parameter values, and the $1\sigma$ uncertainties were obtained at the 16th and 84th percentile levels: $I_{\rm ring}=0.175\pm0.002 ~ \rm mJy~beam^{-1}$, $r_{\rm ring} = 28.59\pm 0.04 ~\rm au$, and $w_{\rm d} = 3.55\pm 0.03~ \rm au$. The best-fit Gaussian profile is shown in the bottom panel of Figure \ref{fig:radialprofile}.

To estimate the intrinsic (deconvolved) width of the ring, we applied the standard convolution relation for Gaussian functions, assuming that both the beam and the intrinsic ring have Gaussian profiles. The deconvolved width $\hat{w}_{d}$ is then given by $\hat{w}_{d} = (w^{2}_{d} - \sigma^{2}_{b})^{0.5}$, where $ \sigma_{b}$ ($=0\farcs048$, or 7.5 au) is the geometric mean of the beam's standard deviations. Applying this relation yields a deconvolved ring width of $\hat{w}_{d} = 1.62\pm 0.04$ au, centered at $r_{\rm ring} = 28.59\pm 0.04 ~\rm au$. We defer discussion of the dynamical implications of this width to Section~\ref{sec:dust_trapping}.

\subsubsection{Dust Trapping}\label{sec:dust_trapping}

The detection of a narrow outer ring structure offers important insights into its formation mechanisms and the underlying physical processes. One plausible explanation is that this substructure results from dust trapping in gas pressure maxima. In these regions, dust grains accumulate at local pressure maxima where their inward radial drift is halted, leading to the formation of confined dust concentrations. This mechanism has been widely proposed to explain the origin of dust ring structures in several disks revealed by ALMA \citep[e.g.,][]{muto_significant_2015, tsukagoshi_flared_2019, yen_kinematical_2020, liu_forming_2024}.

A key diagnostic for assessing the efficiency of dust trapping is the ratio of the dust ring width to the local pressure scale height, $\hat{w}_{\rm d}/h_{\rm p}$. This dimensionless ratio reflects the degree of radial confinement relative to the vertical gas structure and serves as a proxy for the effectiveness of dust accumulation. According to \citet{dullemond_disk_2018}, dust trapping within pressure bumps is expected to produce dust rings narrower than the local pressure scale height.

To determine whether the outer ring in EX~Lupi satisfies this condition, we compute the pressure scale height using the standard relation:

\begin{equation}
h_{\rm p}(r) = \frac{c_s}{\Omega},
\end{equation}

\noindent
where $c_s $ is the local sound speed and $\Omega$ is the Keplerian angular frequency. Substituting the expressions $c_s = \sqrt{k_B T_d(r) / \mu m_p}$ and $\Omega = \sqrt{G M_*/r^3}$, we obtain

\begin{align}\label{eq:scale_height}
h_{p}(r) &= \sqrt{\frac{k_{B} T_d(r) r^{3}}{\mu m_{p} G M_{*}}} \nonumber \\
         &= (0.025\pm0.001) \left(\frac{r}{1~\mathrm{au}} \right)^{1.26}~ \rm au,
\end{align}

\noindent
where $k_{\rm B}$ is the Boltzmann constant, $m_p$ is the proton mass, and $G$ is the gravitational constant. The mean molecular weight is assumed to be $\mu = 2.3$ in atomic units, and we adopt a dynamical stellar mass of $M_{*} = 0.5~M_{\odot}$ \citep{hales_circumstellar_2018}. The disk temperature profile $T_d(r)$, calculated in Section~\ref{sec:disktemp}, is constrained from the observed vertical height and flaring geometry of the disk surface, including the radial dependence of the flaring angle $\varphi(r)$. The uncertainty in the inferred flaring angle propagates directly into the uncertainty of the derived temperature profile $T_d(r)$, and the derivation procedure of $\varphi(r)$ is detailed in Section~\ref{sec:shape_disksurface}.

Taking the dust ring width $\hat{w}_{\rm d} (=1.62\pm0.04 \rm~au)$ at $r_{\rm ring} = 28.59\pm 0.04$ au measured in Section~\ref{sec:outer_ring}, the calculation yields $\hat{w}_{\rm d}/h_{\rm p}(r_{\rm ring}) = 0.95 \pm 0.04$. This ratio is close to unity, suggesting that the radial width of the dust ring is comparable to the local pressure scale height. Since efficient dust trapping is expected only when the ring is substantially narrower than $h_{\rm p}$ \citep{dullemond_disk_2018}, the marginal confinement may indicate moderately effective trapping. 

In addition, the presence of clump-like substructures along the ring may indicate that the underlying gas pressure bump has become unstable, implying that dust trapping is still ongoing but in a dynamically evolving configuration. We discuss this possibility further in Section~\ref{sec:origin_outerring}.

\subsection{Inner Arc}\label{sec:inner_disk}

\begin{figure*}[t]
\centering
\includegraphics[width=0.98 \textwidth]{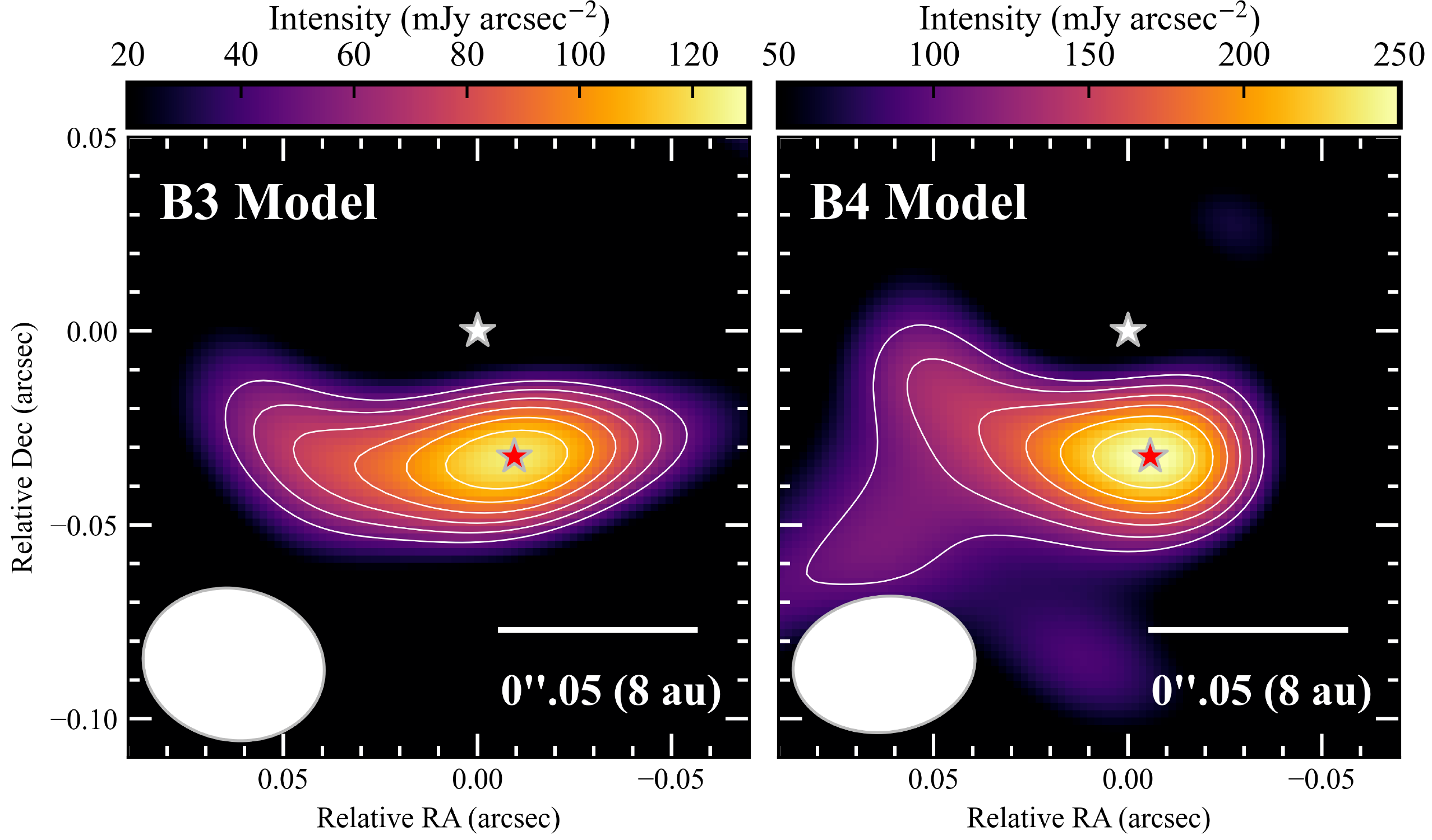}
\caption{Close-up views of the inner arc of EX Lupi in ALMA Band 3 (left) and Band 4 (right) PRIISM model images.  The same linear color scale is used. The white stellar symbol indicates the position of the star and the red one the position of the peak intensity of the inner arc. Contour lines at the $(40,50,60,70,80,90)\%$ value of the peak emission are overplotted. A white bar of $0\farcs05$ (= 8 au) is provided as a reference to the angular scales. The filled white ellipse indicates the effective spatial resolution $\theta_{\rm eff}$.}
\label{fig:innerdisk}
\end{figure*}

Figure~\ref{fig:innerdisk} shows the inner arc structure surrounding the host star, clearly detected in both the Band 3 and Band 4 model images. The existence of an inner disk has previously been inferred from infrared excess in the spectral energy distribution \citep[SED;][]{sipos_ex_2009} and from signatures of stellar accretion \citep{sicilia-aguilar_optical_2012}. This study presents the first direct spatially resolved detections of the inner arc in the dust continuum.

The peak positions of the inner arc in both the model and restored images in Bands 3 and 4 exhibit a consistent upward offset of $30-40$ mas (corresponding to $5-6$ au) relative to the stellar position. To assess the significance of this offset, we use the restored images, which provide well-defined noise levels, and apply the SNR–limited positional accuracy relation given by $ \Delta p / (\mathrm{\rm arcsec})  = 1.1 \times \left\langle\theta \right\rangle /(\text{SNR})$, where $\left\langle\theta\right\rangle$ is the geometric mean of the beam size in the restored image (ALMA Technical Handbook, Equation (10.7)). This calculation yields positional uncertainties of $5-10$ mas in both Band 3 and Band 4 images$-$significantly smaller than the observed offset. The fact that the displacement appears consistently in two independent datasets further confirms that the offset is real and not attributable to noise. This result supports the presence of a spatially offset inner arc surrounding the host star.

In Section \ref{sec:elongation}, we quantify the elongation of the emission of the inner arc. In Section \ref{sec:spectral_index}, we explain the measurement of its spectral index and inferred physical conditions.

\subsubsection{Elongation}\label{sec:elongation}

\begin{figure}[t]
\centering
\includegraphics[width=0.5 \textwidth]{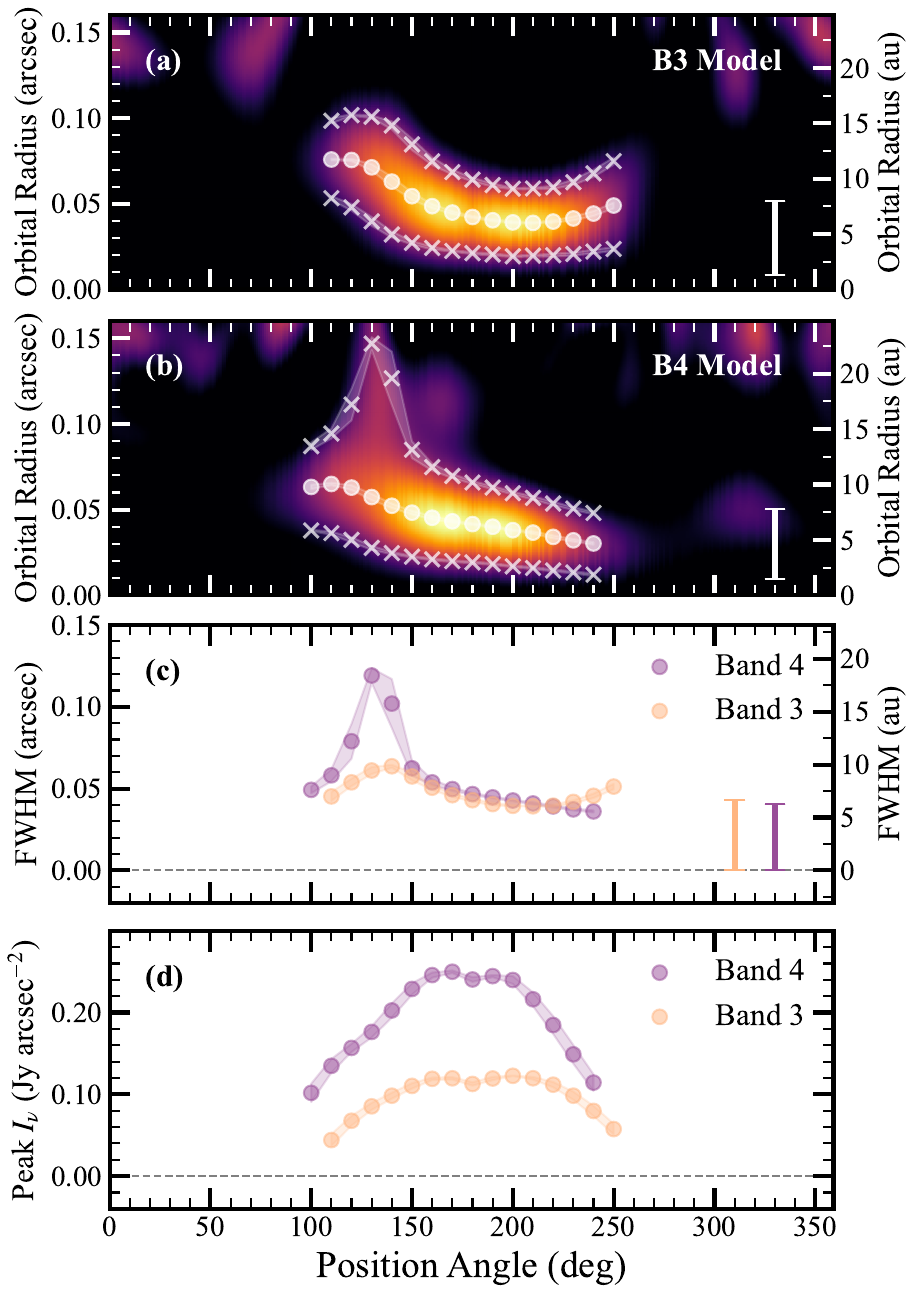}
\caption{Panels of (a) and (b): Deprojected position angle profiles for Band 3 and Band 4 model images. The profiles show the radial location of the peak (circle points) and its $50\%$ of the intensity distribution of the inner arc overlaid by the model images. The origin of the orbital radius is the position of the central star. Panel (c) shows the radial FWHM sizes of the intensity distribution. Panel (d) shows the peak intensity (circle points). In all panels, the circle symbols denote PA-binned means computed over $10^{\circ}$-wide PA bins; the shaded regions indicate the corresponding standard deviations. The vertical bars indicate the geometric mean of the effective spatial resolutions.}
\label{fig:paprofile}
\end{figure}

Figure \ref{fig:paprofile} presents the deprojected PA profiles of the inner arc, derived from the model images. We adopt the same inclination and PA as those of the outer ring (see Section \ref{sec:outer_ring}), under the assumption that the inner arc and the outer ring are aligned on the disk midplane.

By examining the PA profile, we can pinpoint the radial position $r^{\rm peak}_{k}$ of the intensity peak and determine the full width at half-maximum (FWHM) of the intensity distribution within the inner arc. To achieve this, we sample the $k$th data point at intervals of $\Delta \theta = 10^{\circ}$, which provides sufficient data points to trace the radial intensity variations. The width of the ring $w_k$ along the radial direction is then quantified using the standard deviation of the FWHM, defined as FWHM/$2\sqrt{2\ln 2}$.

With these measurements in hand, we can now evaluate the elongation $\varepsilon$ of the inner arc. This elongation is a critical parameter that characterizes the morphology of the ring, and is expressed as the aspect ratio of the arc length to the width of the inner arc:

\begin{equation}
\varepsilon = \frac{\int r d\theta}{\bar{w}} =
\frac{\sum_{k=1}^{N} r^{\rm peak}_k \Delta \theta}{N^{-1}\sum_{k=1}^{N} w_{k}},
\end{equation}
\noindent

Using the derived measurements from the PA profile, we compute the arc length ($\int r d\theta$), which is found to be $\sim 20$ au for both images. Additionally, the averaged ring width $\bar{w}$ is estimated to be at most 3 au, serving as an upper limit since the FWHMs of the intensity distributions are comparable to the spatial resolutions of the images (see Figure \ref{fig:paprofile} (c)).

These calculations yield a lower limit of $\varepsilon > 7$ for the elongation, indicating a significantly elongated structure. Similar elongations have been reported in disks with large, crescent-like rings beyond 50 au, where typical values fall within the range of $\varepsilon = 2-6$ \citep{van_der_marel_major_2021}. The systems analyzed in that study were primarily intermediate-mass stars ($>1.0~M_{\odot}$), reflecting the sample selection available at the time. In this context, the strong elongation observed in EX Lupi is notable given its lower stellar mass ($M_{*}=0.5~M_{\odot}$), suggesting that such pronounced asymmetries are not limited to massive systems.

Our discovery of a small-scale inner arc ($r < 10$ au) connected to an outer ring-gap structure around a low-mass star ($\sim 0.5~M_{\odot}$) presents a striking contrast to these known cases. This finding suggests that the mechanisms driving asymmetric structures may operate across a wider range of disk scales and stellar masses than previously anticipated.

\subsubsection{Spectral Index}\label{sec:spectral_index}

One of the key observational diagnostics for dust growth in disks is the spectral index. By analyzing restored images at two different wavelengths, we can extract crucial information about the dust properties. For the inner arc, which was observed with a relatively high SNR, we calculate the average spectral index using the following equation:
\begin{equation}\label{es:spectral_index}
\begin{split} 
\bar{\alpha}_{\rm mm} = \frac{1}{M \times N} \sum_{i=1}^{M} \sum_{j=1}^{N} \frac{\log \left(I_{\rm B3}(i, j) / I_{\rm B4}(i, j)\right)}{\log \left(\nu_{\rm B3} / \nu_{\rm B4}\right)} \\ \text{subject to } I_{\rm B3} > 3\sigma^{\rm B3}_{\rm noise} \land I_{\rm B4} > 3\sigma^{\rm B4}_{\rm noise},
\end{split} 
\end{equation}
\noindent where $I(i, j)$ is the intensity at pixel indices $i$ and $j$ in a two-dimensional image of the inner arc, and $\nu$ is the observing frequency. The terms $M$ and $N$ represent the total number of pixels along the two axes of $I(i, j)$. The condition ($I_{\nu} > 3\sigma_{\rm noise}$) ensures that pixels contaminated by the dust-tail-like emission$-$detected exclusively in the Band 4 image$-$are excluded from the calculation. To maintain consistency, the restored Band 4 image is smoothed to match the spatial resolution of the Band 3 image.

This calculation yields $\bar{\alpha}_{\rm mm}=2.0\pm0.5$. The uncertainty in $\bar{\alpha}_{\rm mm}$ is estimated by error propagation, taking into account the RMS noise of the images. This value is significantly lower than the typical interstellar medium dust spectral index of \citep[$\alpha_{\rm mm} \sim 3.8$;][]{draine_submillimeter_2006}, suggesting either substantial grain growth within the inner arc under optically thin conditions \citep{soon_investigating_2019} or that the emission is optically thick even in the $90-150$ GHz range. The latter is consistent with recent findings that the bulk emission from protoplanetary disks tends to be dominated by optically thick dust at (sub)millimeter wavelengths \citep{chung_4400_2025}.

\section{Analysis with Archival Near-Infrared Image}
\label{sec:nearinfraded_image}

In this section, we analyze an archival near-infrared scattered-light image of EX Lupi to constrain the geometry of its disk surface. In Section \ref{sec:scatlight_disk}, we examine the azimuthal asymmetry in the scattered light and discuss its origin in relation to the inner arc seen in the ALMA images. In Section \ref{sec:shape_disksurface}, we derive the height and flaring geometry of the disk surface using an elliptical fit to the outer scattered light, and compare the results with theoretical expectations for flared, dust-settled disks.

\subsection{Scattered-light Asymmetry of Disk Surface}\label{sec:scatlight_disk}

\begin{figure*}[t]
\centering
\includegraphics[width=0.95 \textwidth]{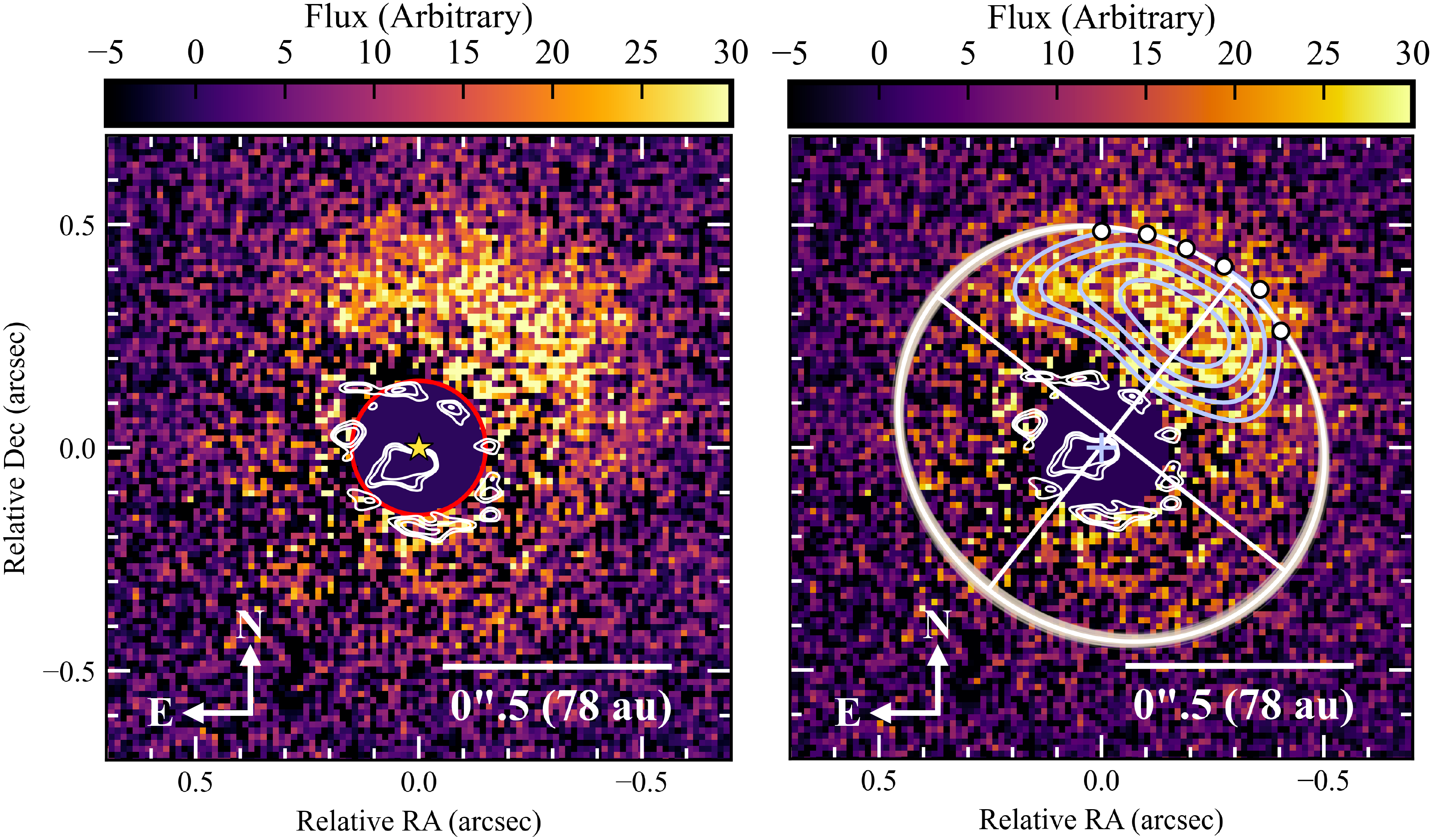}
\caption{The disk emission of EX Lupi observed at near-infrared and millimeter wavelengths. Left: the color map represents the near-infrared polarimetric $Q_{\Phi}$ image, capturing scattered light from micron-sized particles, observed with VLT/SPHERE in the $H$ band \citep[$\lambda = 1.6~\mu$m; ][]{rigliaco_circumstellar_2020}. The red circle marks the inner working angle of the coronagraph with a radius of $0\farcs15$. Overlaid in white contours is the dust continuum emission, tracing millimeter-sized particles, obtained from ALMA Band 4 observations ($\lambda = 2.0$ mm; this work). Contour levels match those in Figure \ref{fig:priism_images}. The yellow star symbol denotes the stellar position.
Right: the best-fit ellipse to the disk edge is displayed as the white ellipse, while the light white curves indicate the distribution of solutions within the estimated $1\sigma$ errors. The blue contours represent the scattered-light image convolved with a Gaussian with a FWHM of $0\farcs2$. Contour levels are set at [60, 70, 80, 90]$\times \sigma_{\rm sca}$, where $\sigma_{\rm sca}$ is the $1\sigma$ uncertainty derived from an emission-free region of the convolved scattered-light image. The white circles along the disk edge indicate the radially averaged positions, measured at azimuthal intervals of $10^{\circ}$ from the convolved scattered light image. The blue cross symbol marks the center of the ellipse model, determined from the outer ring in the ALMA continuum image (see Figure \ref{fig:ellipse_fit}), and is set as the center position of the EX Lupi system.}
\label{fig:sphere_image}
\end{figure*}

\begin{figure*}[t]
\centering
\includegraphics[width=0.95 \textwidth]{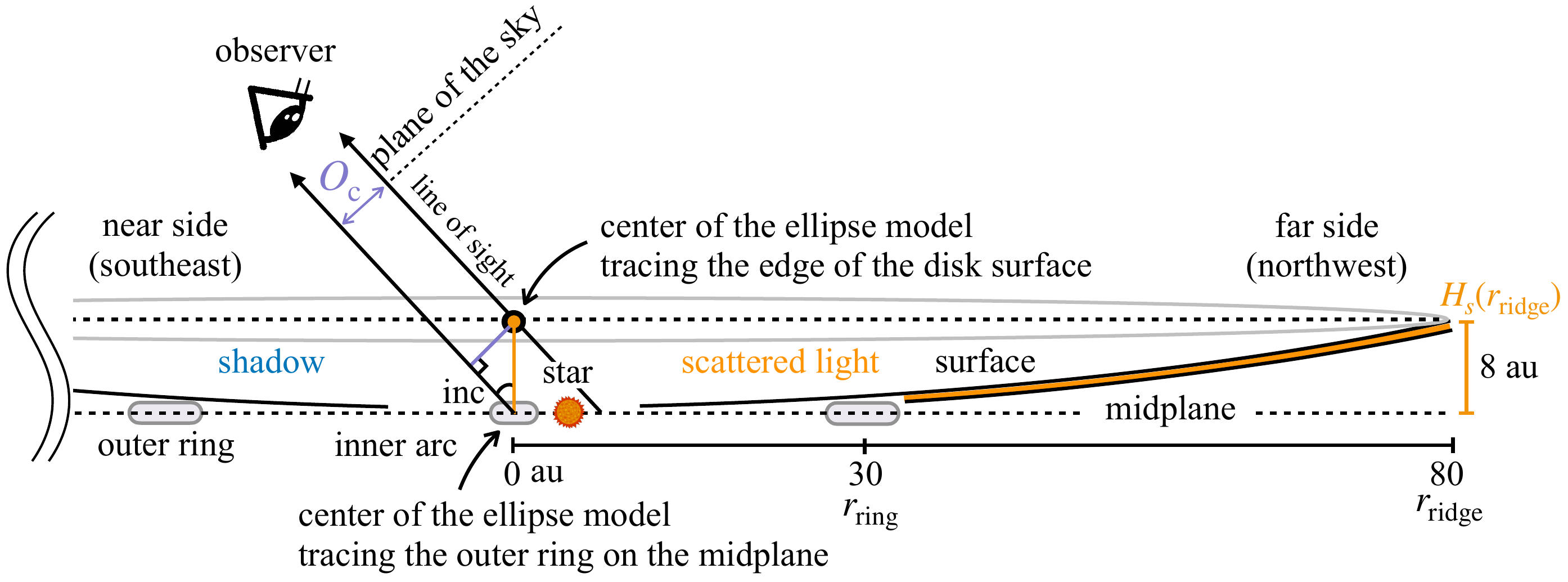}
\caption{Schematic view of the EX Lupi disk structure, based on dust continuum and scattered light imaging observations. The system consists of an inner arc and an outer ring at radius $r_{\rm ring}$ on the midplane. The disk has a height $H_s(r_{\rm ridge})$ of the photospheric surface at radius $r_{\rm ridge}$. Due to the inclination ($\rm inc$) view, the observed disk geometry appears elliptical, with the inclination (inc) angle determined by fitting an ellipse to the projected outer ring on the midplane. The system center is defined to coincide with the center of this ellipse. When viewed at the inclination, the ellipse tracing the edge of the scattering surface appears offset along its minor axis by a measurable displacement $O_{\rm c}$ from the system center, offering the height of the scattering surface. The inner arc plays a critical role in shaping the illumination pattern: it casts a shadow on the near side (southeast) while allowing direct stellar irradiation to reach the far-side (northwest) disk surface, influencing the observed brightness distribution.}
\label{fig:shematicview}
\end{figure*}


Figure~\ref{fig:sphere_image} shows a near-infrared polarimetric $Q_{\Phi}$ image
\footnote[4]{The $Q_{\Phi}$ image represents the azimuthal component of the linearly polarized intensity, computed in a polar coordinate system centered on the star \citep{schmid_limb_2006}. For low-inclination disks dominated by single scattering, the polarization is expected to be oriented azimuthally, making $Q_{\Phi}$ a reliable proxy for polarized intensity \citep{de_boer_polarimetric_2020}. In such cases, $Q_{\Phi}$ effectively traces the disk surface while suppressing noise. However, its interpretation may be affected by multiple scattering or by significant disk inclination. In this study, we use the $Q_{\Phi}$ image primarily as a qualitative tracer of the disk-surface structure, keeping these limitations in mind.}
in the $H$ band ($\lambda = 1.65~\micron$) of EX Lupi obtained with SPHERE at the VLT from \cite{rigliaco_circumstellar_2020}. This observation was conducted in the same year as the ALMA Band 3 and 4 observations presented in Sections~$\ref{sec:observation}-\ref{sec:disk_str_prop}$, providing a contemporaneous view of the disk-surface geometry traced by scattered light. As described by \citet{rigliaco_circumstellar_2020}, the image traces stellar light scattered by submicron-sized dust grains located at the upper layers of the disk atmosphere.

The scattered-light distribution appears strongly asymmetric: The emission is concentrated on the northwest side of the star. This pattern is roughly antialigned with the inner arc detected in the ALMA continuum image (Section~\ref{sec:inner_disk}), which lies southeast of the star. The anticorrelation between the submillimeter continuum and the scattered light suggests that the inner arc casts a shadow onto the outer disk surface, obstructing stellar irradiation in the southeast direction. This suggests that the inner arc is a geometrically thick.

Analogous cases of shadowing due to an inclined or misaligned inner arc have been observed in other systems \citep[e.g.,][]{benisty_shadows_2018, muro-arena_shadowing_2020, bohn_probing_2022}, supporting the idea that localized disk structures can modulate the illumination pattern across the disk surface. In the case of EX Lupi, \citet{hales_circumstellar_2018} reported a blue-shifted molecular outflow extending toward the northwest. Assuming the outflow axis is approximately perpendicular to the disk plane, this geometry implies that the northwest side corresponds to the far side of the disk, from which the approaching outflow is launched. Although forward scattering typically enhances the brightness of the near side \citep[e.g.,][]{takami_surface_2014, ginski_observed_2023}, the near-side region in EX Lupi is likely obscured by the inner arc, which casts a broad shadow across the disk surface. As a result, while the brighter forward-scattered emission from the near side is suppressed, only the fainter scattered light from the far side may remain visible in the scattered-light image.

Taken together, the observed asymmetry in the scattered light may not reflect an intrinsically asymmetric surface density distribution, but instead arise from a geometrically symmetric disk whose illumination is modulated by shadowing from the inner arc.

\subsection{Shape of Disk Surface}\label{sec:shape_disksurface}

The height of the disk surface from the midplane can be geometrically inferred from the scattered-light morphology under the assumption that the disk is circular and azimuthally symmetric. This assumption is commonly adopted in surface geometry reconstructions, particularly when the scattered-light features do not exhibit clear signs of eccentricity \citep{de_boer_multiple_2016, ginski_direct_2016, avenhaus_disks_2018}. When viewed at the moderate inclination, a center of the ellipse tracing the edge of the disk surface can appear to be offset along its minor axis from the center position of the system \citep{ginski_sphere_2024}. Therefore, the offset of the ellipse can be used to calculate the height of the scattering surface of the disk.

We here use the northwest side of the observed near-infrared emission to extract the edge of the disk surface in the scattered light. Appendix \ref{appendix:ellipse_fit} details how the ellipse fits the edge of the disk surface.  The coordinate center for this fitting is set to the center of the elliptical model, defined by fitting the outer ring in the ALMA Band 4 continuum image rather than the stellar position, which is offset from the ellipse center. 

The right of Figure~\ref{fig:ellipse_fit} shows the best-fit result of the ellipse model tracing the edge of the scattered-light disk surface. The ellipse center is offset from the coordinate center by $5.6 \pm 1.6$~au, with the offset being along the minor axis of the elliptical model. This offset $O_{\mathrm{c}}$ provides the disk-surface height $H_{s}(r)$ using a simple trigonometric relation:
\begin{equation}\label{eq:trigonometric_relation}
    H_{s}(r_{\rm ridge}) = \frac{O_{\mathrm{c}}}{\sin(i)},
\end{equation}
where $i~(= 42^{\circ}.6 \pm 2^{\circ}.3)$ represents the disk inclination from the disk midplane as determined from the elliptical fit to the ALMA dust continuum image (see Section \ref{sec:outer_ring}), and $r_{\rm ridge}$ indicates the ridge radius of the disk surface derived from the semi-major axis of the elliptical model. As a reference, Figure \ref{fig:shematicview} presents a geometrical schematic view of the EX Lupi disk to account for the trigonometric relation in Equation \ref{eq:trigonometric_relation}.

This metric finally yields $H_{s}(r_{\rm ridge}) = 8.2 \pm 2.3$~au at $r_{\rm ridge} = 77 \pm 3$~au, corresponding to an aspect ratio of $H_{s} / r = 0.11 \pm 0.03$. The obtained value of $H_{s}$ is consistent with that expected from a theoretical flaring disk model for T Tauri stars, which includes the effect of dust settling \citep{dullemond_effect_2004}. Compared to the empirical power-law relation $H_{s} \propto r^{1.22 \pm 0.03}$ derived from a fit to $H_{s}/r$ measurements of five other protoplanetary disks \citep{avenhaus_disks_2018}, our result is about a factor of 2 lower than the corresponding value of $H_{s}(77~\mathrm{au}) = 15.7 \pm 0.5$~au.

We assume that the radial aspect ratio $H_{s}(r)/r$ of the EX Lupi disk follows a power law with the flaring index $\beta (= 1.09)$ determined from a SED fit using a flared disk model \citep{sipos_ex_2009}. The adopted $\beta$ is close to the theoretical predictions of flared disk models \citep[$\beta = 1.1-1.3$;][]{kenyon_spectral_1987, chiang_spectral_1997}. The disk-surface height $H_0$ at a fiducial radius $r_0$ is then given by 
\begin{equation}\label{eq:aspect_ratio}
\frac{H_{\mathrm{s}}(r)}{r}=\left(\frac{H_{0}}{r_{0}}\right)\left(\frac{r}{r_{0}}\right)^{\beta-1}.
\end{equation}
\noindent
This analysis derives $H_0 = 0.07 \pm 0.02$ au at $r_0 = 1$ au. With $H_{s}(r)/r$, we can derive the flaring angle $\varphi(r)$, which defines the angle at which stellar radiation strikes the disk surface \citep{kusaka_growth_1970, chiang_spectral_1997}:

\begin{equation}\label{eq:flaring_angle}
\varphi(r)= \frac{4}{3 \pi }\frac{R_{\star}}{r}+r \frac{d} {dr}\left \{\frac{H_{\mathrm{s}}(r)}{r}\right\} \simeq r \frac{d} {dr}\left \{\frac{H_{\mathrm{s}}(r)}{r}\right\}, 
\end{equation}

\noindent
where the first term in this equation represents the direct irradiation of the disk surface by stellar radiation (with stellar radius $R_{*}$), while the second term describes the geometrical shape of the disk surface (i.e., its flaring structure). Since $R_{*}\ll r$ at au-scale distances in the scope of our study, the first term becomes negligible. By substituting Equation \ref{eq:aspect_ratio} into Equation \ref{eq:flaring_angle}, we obtain

\begin{align} \label{eq:flaring_angle_second}
\varphi(r) & = H_{0} \cdot \frac{\beta-1}{r_{0}} \cdot\left(\frac{r}{r_{0}}\right)^{\beta-1} \nonumber \\
           & = 0.006 \pm 0.002 \left(\frac{r}{1~\mathrm{au}} \right)^{0.09}.
\end{align}

\noindent
The derived flaring angle at the outer disk edge $r_{\rm edge} (=77 \pm 3 ~\mathrm{au})$ is $0.009\pm0.003$, indicating a subdued flaring structure compared to theoretical models for passively irradiated disks \citep[$\varphi\sim 0.05$;][]{chiang_spectral_1997}. 

To the best of our knowledge, this is the first case where $\varphi(r)$ has been geometrically inferred by combining constraints from scattered light and (sub)millimeter observations, without requiring complex radiative transfer modeling, assuming that the flaring index $\beta$ is externally constrained.

\section{Characterizing the Thermal Structure and Size of the Disk}\label{sec:disk}

To understand the physical conditions of the EX Lupi disk, we characterize its thermal structure and spatial extent. In Section~\ref{sec:disktemp}, we derive the radial temperature profile under the assumption of a passively irradiated flared disk, constrained by the flaring angle, which is estimated using the scattered-light geometry together with the flaring index derived from SED modeling. In Section~\ref{sec:disksize}, we estimate the radial extent of the dust disk using the PRIISM image, and compare it with the gaseous and photospheric components traced by $^{12}$CO line emission and near-infrared scattered light, respectively.

\subsection{Thermal Structure of the Disk}\label{sec:disktemp}
We assume that stellar radiation is the dominant source of disk heating and adopt a passively heated, irradiated flaring disk model without  grain scattering \citep{chiang_spectral_1997, okuzumi_global_2022}. In this framework, the disk surface is inclined at a flaring angle $\varphi(r)$, defined in Equation \ref{eq:flaring_angle_second},
which governs the efficiency of stellar irradiation. The temperature at each radius $r$ is determined by balancing the stellar irradiation, given by $F_{\rm in} = \varphi(r) L_{*} / (8 \pi r^{2})$, with the thermal emission from dust grains, expressed as $F_{\rm out} = \sigma_{\rm SB} T(r)^4$. Solving this energy balance yields
\begin{align}
T_d(r) & = \left\{ \frac{ L_{*} \varphi(r)}{8 \pi r^{2} \sigma_{\rm SB}} \right\}^{1/4} \nonumber \\
       & = (92 \pm 8) \left( \frac{L_{*}}{L_{\odot}} \right)^{0.25} \left( \frac{r}{1~\mathrm{au}} \right)^{-0.48} ~ \mathrm{K},
\end{align}
\noindent
where $\sigma_{\rm SB}$ is the Stefan$-$Boltzmann constant, and $L_{*}$ is the stellar luminosity. When adopting the luminosity in the quiescent phase of $L_{*} = 0.43~L_{\odot}$ \citep{Wang2023ApJ}, the disk temperature profile simplifies to 
\begin{equation}\label{eq:disktemp_simple}
T_d(r) = (75 \pm 6) \left(\frac{r}{1~\mathrm{au}} \right)^{-0.48}. 
\end{equation}
We here define the coordinate origin ($r=0$ au) as the center of the elliptical fit to the outer ring of the EX Lupi system. For simplicity, we do not apply any correction for the offset between this reference center and the actual stellar position (i.e., $r \pm \Delta r$), which deviates by a few au. This approximation introduces only a minor impact on our temperature estimates; at the location of the outer ring ($r=30$ au), the resulting temperature deviation remains within $10~\%$ of the estimated value.

We now compute the disk$-$averaged temperature, defined as $\langle T_{\mathrm{d}} \rangle = \int T_d(r) r \, dr/\int r \, dr, \quad \text{where } r \in [1, 34]~\mathrm{au}$, corresponding to the estimated dust disk size as determined in Section \ref{sec:disksize}. The resulting average disk temperatures yields $18\pm1$ K, which is consistent with the disk$-$averaged temperature ($\langle T_{\mathrm{d}} \rangle  \approx 20$ K) predicted by stellar luminosity scaling relations derived from T Tauri stars \citep{andrews_mass_2013, van_der_plas_dust_2016}.

\subsection{Radial Extent of the Dust Disk}\label{sec:disksize}

Dust emission at (sub)millimeter wavelengths traces pebble$-$sized particles that are concentrated in the disk midplane \citep[e.g.,][]{tsukagoshi_alma_2022, chung_sma_2024}. In this section, we measure the dust disk radius, $r_{\rm dust}$, from the PRIISM model image at Band 4 using the curve-of-growth method, following the approach of \cite{yamaguchi_alma_2024}. First, we deproject the dust continuum image to obtain a face-on view, using the inclination and PA of the dust disk. We then define the incremental flux density based on the radial surface brightness distribution:

\begin{align}
F_{\nu}(r) &= 2 \pi \int_{0}^{r} I_{\nu}\left(r^{\prime}\right)  r^{\prime} \mathrm{d} r^{\prime} \nonumber \\
&= 2\pi \sum_{j=1} I_{\nu}(r_j) r_j \Delta r,
\end{align}

\noindent
where $F_{\nu}(r)$ represents the flux density enclosed within radius $r$, $r_j$ denotes the discrete radial bins, $I_{\nu}(r_j)$ is the surface brightness at radius $r_j$, and $\Delta r$ is the width of each radial bin. In practice, we measure $F_{\nu}(r)$ using successively larger photometric apertures and identify the radius at which $F_{\nu}(r)$ asymptotically approaches a constant value of 5.3 mJy. The dust disk radius $r_{\rm dust}$ is then defined as the radius where $0.95 \times F_{\nu} = F_{\nu}(r)$.

To estimate the uncertainty $\sigma_r$ in the disk radius, we use the effective spatial resolution and assume $\sigma_{\mathrm{r}}=\left\langle\theta_{\mathrm{eff}}\right\rangle /2 \sqrt{2 \ln 2}$, where $\left\langle\theta_{\rm eff}\right\rangle$ denotes the geometric mean of the spatial resolution. This calculation yields a dust disk radius of $r_{\rm dust} = 0\farcs22\pm 0\farcs02$ ($34\pm3$ au).

The derived $r_{\rm dust}$ suggests a compact dust disk, consistent with the known properties of most protoplanetary disks around low-mass M- to K-type stars. These disks are typically faint and compact, with total fluxes below 100 mJy at (sub)millimeter wavelengths and dust radii smaller than 40 au \citep{yamaguchi_alma_2024, shoshi_alma_2025}. The measured radius also aligns well with compact dust radii observed in other EXor systems \citep{cieza_alma_2018}.  

In contrast, \cite{hales_circumstellar_2018} find that optically thick $^{12}$CO line emission reveals a significantly larger gaseous disk around EX Lupi, extending up to 150 au in radius. This finding aligns with theoretical predictions and observations suggesting that millimeter-sized grains undergo inward radial drift due to gas drag, leading to a more concentrated dust distribution \citep{weidenschilling_aerodynamics_1977, testi_dust_2014}. 

The photospheric surface of the disk extends to an intermediate radius of $r_{\rm ridge} = 77 \pm 3$ au (see section \ref{sec:shape_disksurface}). \cite{rigliaco_circumstellar_2020} present an $r^{2}$-scaled scattered-light image, corrected for stellar illumination dilution, which shows no further extended disk-surface feature. This suggests that small (micron-sized) grains are coupled to the gas but do not extend to the full radius of the $^{12}$CO disk, or it could be due to a self-shadowing geometry where the inner region of the disk intercepts and reprocesses a large fraction of the stellar photons \citep{dullemond_effect_2004, garufi_sphere_2022}. If the latter is true, then the true radius of the disk surface is larger than what we have measured.

\section{Discussion} \label{sec:discussion}

In this section, we summarize the observational results and consider their implications for the physical structure, evolution, and possible dynamical drivers within the EX Lupi disk system. We organize the discussion into four subsections, addressing: (1) the nature and origin of the inner arc, (2) the formation scenario of the outer ring, (3) the global evolutionary status of the disk, and (4) the possible link between disk structure and episodic accretion outbursts.

\subsection{Origin of the Inner Arc}\label{sec:origin_innerarc}

The ALMA images presented in Section~\ref{sec:outer_ring} reveal the inner arc within 10 au of the central star, exhibiting a strong elongation ($\varepsilon > 7$). The low spectral index measured in this region ($\alpha_{\rm mm} = 2.0 \pm 0.5$) may trace moderately grown dust grains, though the evidence remains tentative. Overall, these results support the interpretation of the arc as a physically real and possibly evolved structure.

One plausible origin for this nonaxisymmetric feature is the formation of a vortex via Rossby wave instability, which can be triggered at the edges of a gap carved by a sufficiently massive companion \citep{lovelace_rossby_1999, hammer_slowly-growing_2017}. In such scenarios, pressure gradients generated by the companion induce anticyclonic vortices that act as efficient dust traps, concentrating large grains into localized arcs \citep{regaly_possible_2012, birnstiel_lopsided_2013}. Alternatively, asymmetric dust accumulation may result from tidal gas flows across a companion-induced cavity, leading to azimuthal mass pileup at the cavity edge \citep{ragusa_origin_2017, calcino_signatures_2019}. In both cases, a more massive companion tends to induce stronger disk asymmetries.

We note that previous radial velocity (RV) monitoring of EX Lupi has reported a periodic signal that may be attributed to a close-in giant companion with $m \sin i = 14.7M_{\rm Jup}$ in a 6 days orbit ($r = 0.06$ au) \citep{kospal2014}. Assuming the companion shares the inclination of the disk ($i \sim 42^\circ.6$), this corresponds to a mass of $M_p\sim22~M_{\rm Jup}$ (or $M_p/M_\star \sim 0.044$), placing it firmly in the brown dwarf regime \citep[$13-80~M_{\rm Jup}$;][]{hayashi_evolution_1963, spiegel_deuterium-burning_2011}. Nevertheless, this finding remains tentative. The RV signal itself could alternatively arise from line-dependent veiling associated with accretion processes, rather than orbital motion \citep{sicilia-aguilar_accretion_2015}. Moreover, it remains unclear whether a companion located as close as $r < 0.1$ au can directly give rise to nonaxisymmetric dust structures at radial distances of 3-10 au. This large spatial separation may challenge conventional models of localized disk interaction.

An alternative possibility is that the inner arc is not shaped by the putative close-in companion inferred from RV monitoring, but rather by an as-yet-undetected massive companion orbiting just inside the arc (e.g., at a few au). Such a companion could naturally open a gap and induce asymmetries via vortex formation or tidal gas flows, as discussed above. This scenario would also alleviate the need for long-range dynamical coupling from the close-in companion at $r<0.1$ au, which remains difficult to reconcile with current models.

We note that the detected tail-like structure connected to the inner arc is observed in the Band 4 continuum, implying the presence of millimeter-sized dust grains possibly nearby on the disk midplane. This contrasts with streamer-like features seen in scattered light, which typically trace micron-sized grains in the surface layers of the disk \citep[e.g.,][]{keppler_gap_2020}. The physical origin and grain properties of the tail remain uncertain, and require further observational and modeling efforts for clarification.

\subsection{Origin of the Outer Ring}\label{sec:origin_outerring}

As demonstrated in Section~\ref{sec:dust_trapping}, the narrow outer ring at 30 au exhibits properties indicative of dust trapping at a gas pressure maximum. The formation of such pressure maxima is often attributed to the gravitational interaction between the disk and a massive embedded companion \citep[e.g.,][]{muto_two-dimensional_2010}. As the companion exchanges angular momentum with the surrounding gas, it can carve a gap in the disk and reshape the radial pressure profile. This interaction naturally produces a local pressure maximum at the outer edge of the gap, where millimeter-sized dust grains accumulate \citep{pinilla_astrophysics_2012, pinilla_hints_2020}.

Observationally, our ALMA images (Section~\ref{sec:outer_ring}) reveal a $\sim$5 au offset between the ring center and the star, consistent with disk eccentricity excited by a massive companion \citep{kley_disk_2006, hsieh_secular_2012, tanaka_eccentric_2022}. We also identify clump-like substructures along the ring, suggesting that the pressure bump may no longer be in a long-lived, dynamically stable phase but has begun evolving. Such conditions are prone to Rossby wave instability \citep[e.g.,][]{ono_parametric_2016, ono_parametric_2018}, potentially leading to multiple vortex formation, seen as the clumps in the azimuthal direction. Dust originally trapped in this unstable ring could survive over a relatively wide range of parameter space, including the ring width ($\hat{w}_{\rm d}/h_{\rm p}\sim 1$) and background disk properties relevant to this study \citep{chang_origin_2023}.

If the companion undergoes inward migration, as expected from angular momentum exchange with the disk gas, the dust ring may remain at the location of the original pressure maximum, resulting in a residual ring structure. Two-fluid hydrodynamical simulations of low-viscosity disks (e.g., $\alpha \sim 10^{-4}$) by \citet{kanagawa_dust_2021} show that the dust ring remains at the initial pressure maximum and does not follow the migrating companion. Their models further predict the formation of a second, inner arc near the companion's new position, potentially giving rise to a multiring morphology. When the companion accretes gas efficiently during its migration, its mass can increase significantly \citep{durmann_accretion_2017}. As its mass grows, the dust distribution around it may become increasingly asymmetric, forming a crescent-shaped, incomplete ring rather than a symmetric annulus \citep{kanagawa_dust_2021}. This combined effect of migration and mass growth could naturally account for asymmetric features such as the inner arc observed in this study.

Alternatively, the outer ring may have been sculpted by a second, more distant companion that remains undetected. Hydrodynamical simulations show that in multicompanion systems each body can create its own pressure maximum, leading to the formation of multiple ring-gap structures through dust trapping \citep[e.g.,][]{pinilla_gas_2015}. If such a configuration applies to EX Lupi, the outer ring could reflect dust accumulation induced by a separate dynamical influence.

\subsection{Implications for Disk Evolution in EX Lupi}

By analyzing both the ALMA dust continuum and the VLT/SPHERE scattered-light images, we infer that the EX Lupi disk exhibits characteristics of a relatively evolved protoplanetary disk. First, the low photospheric surface height ($H_s/r \sim 0.1$ at $r = 77$ au; Section~\ref{sec:shape_disksurface}) suggests that vertical dust settling is already underway.  Second, the disk-averaged temperature (Section~\ref{sec:disktemp}) is comparable to that of typical T Tauri disks, indicating a similar thermal structure.

Furthermore, during its quiescent phase, EX Lupi exhibits mass-accretion rates consistent with those of classical T Tauri stars \citep{lorenzetti_nature_2012} and lacks a residual dust envelope \citep{sipos_ex_2009}, suggesting that the system has already transitioned beyond the embedded phase. The detection of spatially resolved substructures, namely the narrow outer ring and the inner arc, further supports the notion that the disk is thermally and structurally mature and is actively undergoing dust evolution and potential dynamical perturbations.

We also emphasize that the multiwavelength observations (from near-infrared to millimeter) have revealed a complex interplay between different dust populations. Given the wavelength dependence of dust emission and scattering \citep[e.g.,][]{Benisty2023}, these differences are not contradictory, but rather reflect how grains of different sizes respond to the gas: small grains remain well coupled to the gas and trace smooth, flared structures, while large grains decouple and concentrate into compact features on the disk midplane. This coexistence of geometrically flared small dust and concentrated large dust represents a property of evolved disks \citep{avenhaus_disks_2018, ginski_sphere_2024}.

Taken together, these observations suggest that the EX Lupi system, although classified as an EXor, harbors a dust disk whose physical and morphological properties are comparable to those of more quiescent T Tauri stars. This highlights the possibility that episodic accretion phenomena may occur even in mature disk environments.

\subsection{Episodic Accretion Outbursts Induced by a Massive Companion}

Theoretical and observational studies suggest that a massive companion can trigger episodic accretion outbursts in young stellar objects \citep{lodato_massive_2004, nayakshin_youngest_2024}. These interactions can lead to the accumulation of material in the disk, followed by its sudden release onto the star, thereby driving accretion outbursts. Given the possibility of the presence of a close-in massive companion in the EX Lupi system, as discussed earlier, it is plausible that similar mechanisms influence the observed variability in accretion rates. The companion could facilitate dust accumulation, leading to the formation of the inner arc. Once a critical threshold in mass is reached, the inner arc may undergo an outburst of enhanced accretion. Such mechanisms provide a compelling explanation for the episodic nature of EX Lupi's accretion history. Future monitoring of accretion variability, coupled with detailed modeling of disk--companion interactions, will be essential to further elucidate these processes.

\section{Conclusion} \label{sec:conclusion}

We present ALMA Band 3 and 4 observations of the EXor-type young star EX Lupi, reconstructed using the PRIISM super-resolution imaging technique based on sparse modeling. Compared to the conventional CLEAN method, PRIISM achieves superior image fidelity and spatial resolution, yielding effective beam sizes of $47\times39$ mas in Band 3 and $47\times35$ mas in Band 4, corresponding to improvements in resolution by factors of 2.1 and 1.3, respectively. While both CLEAN model and PRIISM reproduce the observed visibility data with reduced chi-square values close to unity, PRIISM mitigates the patchy artifacts typical of CLEAN and reconstructs smoother, physically plausible disk morphologies. The key physical findings with respect to the EX Lupi disk, as revealed by the PRIISM images are summarized below:

\begin{enumerate}

\item{We detect two previously unresolved substructures in the EX Lupi disk: a narrow outer ring at $r = 28.6 \pm 0.04$ au, and an asymmetric inner arc within 10 au of the star. This study presents the first spatially resolved detection of both inner and outer substructures in dust continuum emission around EX Lupi, revealing a multiscale disk morphology in an EXor disk.}

\item{The inner arc exhibits a crescent-shaped asymmetric emission pattern with an elongation factor $\varepsilon > 7$, and a peak offset of $5-6$ au from the stellar position. The millimeter spectral index in this inner arc is $\bar{\alpha}_{\rm mm} = 2.0 \pm 0.5$, suggesting either dust grain growth or optically thick emission in the $90-150$ GHz range. The asymmetric distribution of near-infrared scattered light observed using VLT/SPHERE is consistent with the idea that the inner arc casts a shadow on the disk surface.}

\item{The outer ring has a deconvolved radial width of $\hat{w}_{\rm d} = 1.62 \pm 0.04$ au, corresponding to a width-to-scale height ratio of $\hat{w}_{\rm d}/h_{\rm p} = 0.95 \pm 0.04$ at a ring radius of $28.6$ au. This ratio is close to unity, indicating that the dust ring is only marginally narrower than the local gas pressure scale height. Such a configuration suggests moderately effective dust trapping, consistent with dust concentration due to radial drift. The presence of localized clumpy substructures may further imply that the trapping site is not dynamically stable but starts evolving.}

\item{Geometric analysis of the VLT/SPHERE scattered-light image, in combination with the ALMA continuum geometry, reveals a disk-surface height of $H_s = 8.2 \pm 2.3$~au at $r = 77$~au, corresponding to an aspect ratio of $H_s/r = 0.11 \pm 0.03$. Assuming a fixed flaring index of $\beta = 1.09$, this yields a flaring angle of $\varphi(r) = 0.006\pm0.002 \left( r / 1~\mathrm{au} \right)^{0.09}$. The relatively small flaring angle ($\varphi \sim 0.01$ at 77~au) suggests that the disk is moderately flared and vertically settled. Using this flaring geometry under the assumption of radiative equilibrium, we derive a disk temperature profile with a disk-averaged temperature of $18 \pm 1$~K. This value is consistent with those of typical T~Tauri disks, indicating a similar thermal structure.}

\item{The coexistence of a narrow outer ring and a crescent-shaped inner arc suggests that EX Lupi hosts a dynamically evolving yet structurally mature disk. One plausible scenario is that interactions between the disk and a single massive companion migrating inward first induces the dust trap observed as the outer ring, and subsequently induced the arc in the inner region via vortex formation or tidal flows. Alternatively, the outer ring may have been sculpted by a second, more distant companion. These companion-driven mechanisms offer a unified explanation for the observed substructures, and may also regulate the episodic accretion outbursts seen in EX Lupi through mass accumulation and release in the inner arc.}

\end{enumerate}

As described above, we propose that the inner arc obscures the stellar light in the southeast direction, casting a shadow. Such shadows are known to exhibit temporal variability \citep{stolker_polarized_2017, pinilla_variable_2018, muro-arena_shadowing_2020}. Given the Keplerian clockwise rotation of the EX Lupi disk \citep{hales_alma_2020}, the arc is expected to corotate, implying that the illuminated region on the disk surface should gradually shift clockwise. Future multiepoch observations will provide a critical opportunity to test this prediction and to further explore the dynamic interplay between disk structure and stellar irradiation in young, variable systems such as EX Lupi.

\section*{Acknowledgment}
The authors thank the anonymous referee for their comments and advice that helped improve the manuscript and the contents of this study. M.Y. and M.T. acknowledge support from the National Science and Technology Council (NSTC) of Taiwan (grant Nos. 112-2124-M-001-014, 112-2112-M-001-031, 113-2124-M-001-008, 113-2112-M-001-009, and 114-2112-M-001-002). H.B.L. is supported by the National Science and Technology Council (NSTC) of Taiwan (grant Nos. 111-2112-M-110-022-MY3 and 113-2112-M-110-022-MY3). P.G. is supported by the National Science and Technology Council (NSTC) of Taiwan (grant No. 113-2112-M-001-012). This work was supported by NAOJ ALMA Scientific Research grant code 2022-22B. This paper makes use of the following ALMA data: ADS/JAO.ALMA$\#$2017.1.00388.S. ALMA is a partnership of ESO (representing its member states), NSF (USA) and the NINS (Japan), together with the NRC (Canada), NSTC and ASIAA (Taiwan), and KASI (Republic of Korea), in cooperation with the Republic of Chile. The Joint ALMA Observatory is operated by ESO, AUI/NRAO, and NAOJ.

This work has made use of data from the European Space Agency (ESA) mission
{\it Gaia} (\url{https://www.cosmos.esa.int/gaia}), processed by the {\it Gaia}
Data Processing and Analysis Consortium (DPAC,
\url{https://www.cosmos.esa.int/web/gaia/dpac/consortium}). Funding for the DPAC
has been provided by national institutions, in particular the institutions
participating in the {\it Gaia} Multilateral Agreement.

\software{AnalysisUtilities \citep{hunter_2023}, Astropy \citep{astropy2022}, CASA \citep{CASATeam2022}, emcee \citep{Foreman2013}, matplotlib \citep{Hunter2007}, PRIISM \citep{Nakazato2020, Nakazato_priism_2020}, SciPy \citep{2020SciPy-NMeth}}

\restartappendixnumbering  
\appendix

\section{CLEAN Images}\label{appendix:clean_images}

\begin{figure*}[t]
\centering
\includegraphics[width=0.98 \textwidth]{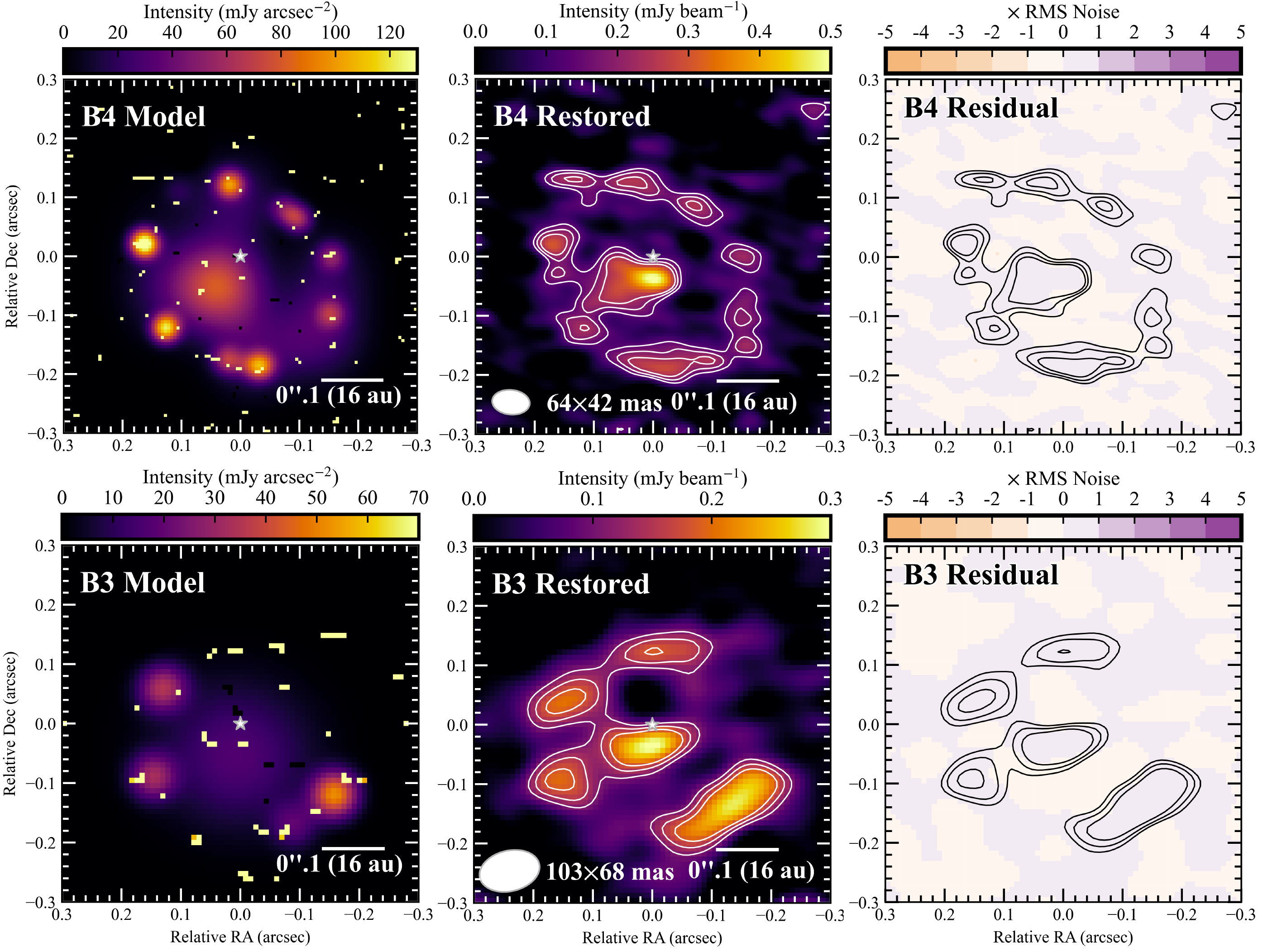}
\caption{Same as Figure \ref{fig:priism_images}, but with the CLEAN procedure (Section \ref{sec:clean_imaging}). Left: CLEAN model generated by  multi scale approach with scale parameters of $[0,1,3]\times\theta_{\rm cl}$, where $\theta_{\rm cl}$ denotes the CLEAN beam size. The image initially has a unit of janskys per pixel ($\rm Jy~pixel^{-1}$), while it is converted to millijanskys per square arcsecond ($\rm mJy~arcsec^{-2}$). Middle: Restored CLEAN image after convolution with an elliptical Gaussian representing the main lobe of the synthesized beam (i.e., the CLEAN beam), along with adding the dirty map of residuals (right). Each white contour corresponds to $[3, 4, 5]\times \sigma_{\rm noise}$, where $\sigma_{\rm noise}$ is the rms noise value calculated from the emission-free area ($36~\mu \rm Jy~\rm beam^{-1}$ for Band 3 and $46~\mu \rm Jy~\rm beam^{-1}$ for Band 4). Right: the residual map. The unit is expressed in rms noise $\sigma$. The black contours represent the same as the white ones in the restored CLEAN image.}
\label{fig:clean_images}
\end{figure*}

Figure~\ref{fig:clean_images} presents a gallery of CLEAN images and their corresponding residual maps at Bands 3 and 4, produced using a Briggs robustness parameter of 0.5. The spatial distributions of the dust continuum in the restored CLEAN images are broadly consistent with those obtained using PRIISM (Figure~\ref{fig:priism_images}). However, the CLEAN model images show noticeable discrepancies compared to the PRIISM model images, reflecting the fundamentally different assumptions underlying each imaging method.

\section{Performance of PRIISM Imaging}\label{sec:priism_image}

In this section, we assess the performance of the PRIISM imaging technique through a series of quantitative evaluations. Appendix~\ref{sec:goodnessfit} evaluates the agreement between the reconstructed images and the observed visibilities via reduced chi-squared statistics in the visibility domain. In Appendix~\ref{sec:effective_resolution}, we estimate the effective spatial resolution of the PRIISM images using a point-source injection experiment. Appendix~\ref{sec:restored_image} describes the generation of restored images by applying beam convolution and residual addition, which are used for analyses requiring accurate noise characterization.

\subsection{Goodness of Fit to Observations}\label{sec:goodnessfit}

\begin{figure*}[t]
\centering
\includegraphics[width=0.95 \textwidth]{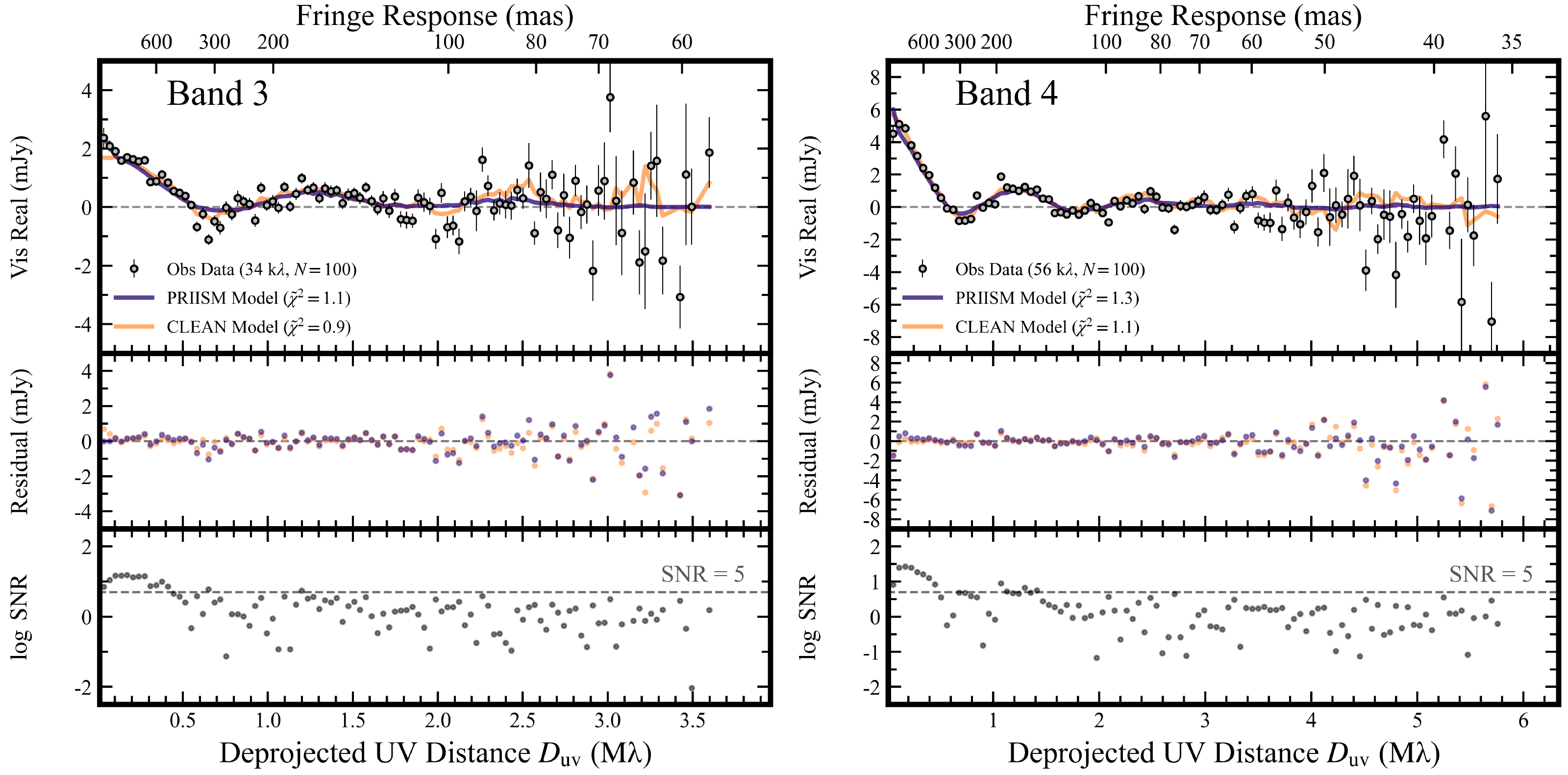}
\caption{Azimuthally averaged radial visibility profiles. The left and right sides correspond to Band 3 and 4 data, respectively. The panels show real parts of the visibilities for the PRIISM image (purple) and the CLEAN model image (orange), together with the observational data (gray). Each profile is binned to 100 data points in total and deprojected to compensate for the disk orientation. The reduced chi-squared values $\tilde{\chi}^{2}$ calculated from the observed data and the models are shown in the labels of the top panels. The panels display, from top to bottom, the amplitude of the real part of the visibility, the residuals obtained by subtracting the model from the observations, and the SNR of the visibility within each bin in $uv$ space (i.e., the ratio of visibility amplitude to the error obtained from the error propagation of the visibility weight). The more detailed calculation of each panel is described in \cite{yamaguchi_alma_2024}.}
\label{fig:vis_profile}
\end{figure*}

To evaluate the goodness of fit between the models (PRIISM, with CLEAN as a comparison) and the observations, we calculated the reduced-chi-squared $\tilde{\chi}^{2}$ using the azimuthally averaged visibility profiles (see Appendix 4 in \citealp{yamaguchi_alma_2024} for details). Figure \ref{fig:vis_profile} shows that the PRIISM model images yield $\tilde{\chi}^{2}$ values around unity, similar to the CLEAN models, meaning both models agree with the observations in the visibility domain.

The situation changes when we consider the image domain. The CLEAN model image exhibits a patchy pattern (see Figure \ref{fig:clean_images}), which can be attributed to the underlying assumption that the multi-scale CLEAN represents the disk structure as a sum of emission components at different size scales but with a mismatch to the ground truth \citep{cornwell_multiscale_2008}. This assumption itself is not enough to adequately model the expected smooth disk structures. The PRIISM image reconstructs smoother structures than the CLEAN model and therefore seems more reasonable. This is because, with the adoption of multiresolution regularization using TSV, a precise smooth image that maintains consistency with observations can be reconstructed using PRIISM.

\subsection{Effective Spatial Resolution}\label{sec:effective_resolution} 

We evaluate the effective spatial resolution $\theta_{\rm eff}$ of each PRIISM model image using the point-source injection method outlined in \cite{yamaguchi_alma_2021}. We inject an artificial point source into the observed visibility data. The flux density of the point source is set to $20\%$ and $10\%$ of the total flux density of the target for Band 3 and 4, respectively. The artificial point source is placed in an emission-free area but at a distance within the MRS. We test the injection of the point source by changing directions along the east, west, north, and south sides, and independently perform the PRIISM imaging. The imaging is carried out on the point-source-injected data using the same set of regularization parameters employed to generate the optimal image of the noninjected data. Each injected point source shows a Gaussian-like distribution in the reconstructed image. We fit it with an elliptical Gaussian function to obtain the FWHM, which measures the effective spatial resolution.

The averaged effective spatial resolutions, $\theta_{\rm eff}$, are  ($47\pm4 \times 39\pm1$ mas, PA = $78^{\circ}\pm 6^{\circ}$) for Band 3 and  ($47\pm1 \times 35\pm2$ mas, PA = $97^{\circ}\pm 13^{\circ}$) for Band 4. The errors indicate the standard deviation of the measurements. The measured spatial resolutions change by only a few percent when the point source is injected in different directions, and the total flux density of the point source is recovered within a $10\%$ error range. Here, we employ the averages as the effective spatial resolutions  $\theta_{\rm eff}$. The effective spatial resolution improved by factors of 2.1 (Band 3) and 1.3 (Band 4) relative to the CLEAN images.

\subsection{Restored Image}\label{sec:restored_image}

Following a customary procedure, we estimated the intensity uncertainties on the PRIISM continuum image\footnote[3]{We note that estimating uncertainties in the model image is not straightforward. PRIISM model images exhibit unexpected positive emissions outside their target source areas due to thermal and systematic noise \citep{yamaguchi_alma_2024}. This is attributed to the adoption of nonnegative constraints in the imaging algorithm, resulting in artificial emissions with positive intensity in the off-source area.} by producing a ``restored'' image $I_{r}$, also adopted in the CLEAN algorithm \citep[e.g.,][]{thompson_interferometry_2017} and another nonparametric imaging with a maximum entropy method \citep[][]{casassus_compact_2015, casassus_variable_2022}. A model image $I_{m}$ is restored through convolution, with an elliptical Gaussian $B$ representing the main lobe of the synthesized beam (i.e., the CLEAN beam), along with the addition of the dirty map of residuals, $I_{d}$, the inverse Fourier transform of the gridded visibility residuals. This restoration procedure (i.e., $I_{r} = I_{m}*B +I_{d}$) was executed with a Briggs weighting of 0.5. The final restored images are expressed in units of $\rm Jy~beam^{-1}$, with a beam size of $92 \times 66 ~ \rm mas ~ (\rm PA = -84^{\circ})$ for Band 3 and $60 \times 38 ~ \rm mas ~ (\rm PA = 87^{\circ})$ for Band 4. The noise levels for Band 3 ($\sigma^{\rm B3}_{\rm noise}$) and Band 4 ($\sigma^{\rm B4}_{\rm noise}$) were calculated from the emission-free area and derived as $37~\mu \rm Jy~\rm beam^{-1}$ and $48~\mu \rm Jy~\rm beam^{-1}$, respectively. These noise levels and SNR are comparable to those of the restored CLEAN images.

We note that the beam convolution process used to generate the restored images leads to a degradation in spatial resolution compared to the original PRIISM model images. Specifically, the effective spatial resolution is reduced by $40\%$ in Band 3 and $20\%$ in Band 4 after convolution. To balance resolution and noise characterization, we adopt a dual-image strategy: the original PRIISM model images are used for a structural analysis, where maximizing spatial resolution is essential; in contrast, the restored images, which incorporate both beam convolution, are used for analyses sensitive to image noise such as in spectral index estimation.

\section{Elliptical Fitting Procedure}\label{appendix:ellipse_fit}

\begin{figure*}[t]
\centering
\includegraphics[width=0.9 \textwidth]{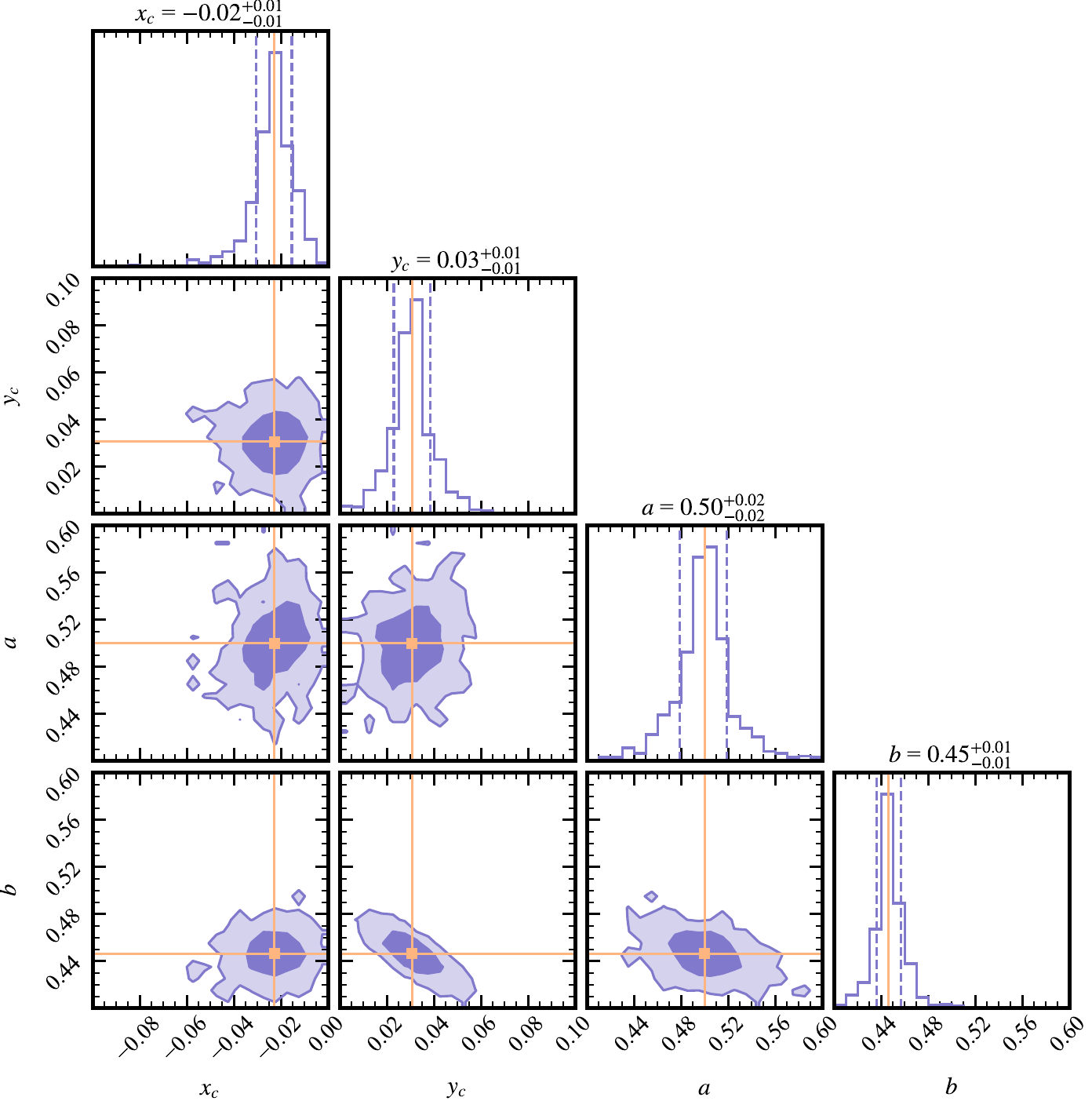}
\caption{Posterior distributions of the best-fit ellipse model parameters, which trace the edge of the scattered EX Lupi disk surface as viewed along the line of sight. All parameters are expressed in units of arcseconds. The purple vertical dashed lines indicate the 16th and 84th percentiles, while the orange solid lines mark the most probable values (50th percentile). Contours represent the $68\%$ and $95\%$ confidence regions. The axis limits of each subplot correspond to the parameter range defined by the uniform prior.
}\label{fig:exlup_cornerplot}
\end{figure*}

As described in Section~\ref{sec:shape_disksurface}, we fit the outer edge of the observed near-infrared scattered light (Figure~\ref{fig:sphere_image}) with an ellipse in order to measure the offset $O_{\rm c}$ of the ellipse center from the coordinate origin. This offset is used to estimate the height of the disk surface above the midplane. The original scattered-light image was first convolved using a Gaussian with a FWHM of $0\farcs2$ to enhance the edge of the disk surface (see blue contours in Figure~\ref{fig:sphere_image}). The convolved image $I_{\rm sca}$ was then transformed into an edge-enhanced image using the Sobel operator implemented in $\tt scikit-image$ \citep{Walt2014PeerJ}, which approximates the image gradient as
\begin{equation}\label{eq:edge_image}
|\nabla I_{\rm sca}(x, y)|=\sqrt{\left|\frac{\partial I_{\rm sca}}{\partial x}\right|^{2}+\left|\frac{\partial I_{\rm sca}}{\partial y}\right|^{2}},
\end{equation}
\noindent
where ($x$, $y$) are the pixel coordinates. To extract the edge structure, we converted the gradient image into a polar coordinate system centered on the star and sampled radial intensity profiles at every $1^{\circ}$ azimuthal angle $\theta$. The radial position of the local maximum (i.e., edge) $r_{\rm peak}(\theta)$ was recorded. In the bright sector between $300^{\circ}$ and $360^{\circ}$, we further computed the local standard deviation $s(\theta)$ over a $\Delta \theta = 10^\circ$ azimuthal window to quantify the positional uncertainty. The resulting $(r_{\rm peak}(\theta), \theta)$ points were converted into Cartesian coordinates $(x_{\rm obs}, y_{\rm obs})$ for use in subsequent ellipse fitting.

The ellipse is here parameterized by its center $(x_c, y_c)$, semi-major axis $a$, semi-minor axis $b$, and position angle $PA$. To reduce parameter degeneracy, we fixed the $PA$ to $51^\circ.7$, as determined from the ALMA continuum image (see Section \ref{sec:outer_ring}). The model predicts the location of a point on the ellipse as a function of the angular parameter $\phi$:

\begin{align}
x_e(\phi) &= x_c + a \cos\phi \cos(PA) - b \sin\phi \sin(PA), \notag \\
y_e(\phi) &= y_c + a \cos\phi \sin(PA) + b \sin\phi \cos(PA).
\end{align}
\noindent
For each observed point $(x_{\rm obs}, y_{\rm obs})$, we numerically minimize the Euclidean distance to the ellipse:

\begin{equation}
\begin{aligned}
&\phi_{\mathrm{opt}} = \arg \min _{\phi \in [0, 2\pi]} d(\phi) \\
                    &\mathrm{with} \quad d(\phi) = \sqrt{\left(x_{e} - x_{\mathrm{obs}}\right)^{2} + \left(y_{e} - y_{\mathrm{obs}}\right)^{2}},
\end{aligned}
\end{equation}
\noindent 
where the optimization was performed using the \texttt{scipy.optimize.minimize} routine with the Nelder-Mead method \citep{2020SciPy-NMeth}. The minimum distances $d_i = d(\phi_{\mathrm{opt}, i})$ across all $N$ observed points are then used to construct the Gaussian log-likelihood function:

\begin{equation}
\ln \mathcal{L}(\mathbf{p}) =  -\frac{1}{2} \sum_{i} \left[ \frac{d_i^2}{s_i^2} + \ln(2\pi s_i^2) \right],
\end{equation}
where $\mathbf{p} = (x_c, y_c, a, b)$ are the free parameters, and $s_i = s(\theta_i)$ is the standard deviation measured locally at each angle $\theta_i$. To incorporate prior knowledge about the parameters, a uniform prior is applied with the following ranges: $x_c \in [-0.1, 0.0]$ arcsec, $y_c \in [0.0, 0.1]$ arcsec, $a \in [0.4, 0.6]$ arcsec, and $b \in [0.4, 0.6]$ arcsec. The prior is defined as

\begin{equation}
\ln \mathcal{P}(\mathbf{p}) =
\begin{cases}
0 & \text{if } p_j \in [p_j^{\text{min}}, p_j^{\text{max}}] \, \forall j, \\
-\infty & \text{otherwise}.
\end{cases}
\end{equation}

The posterior probability is expressed as
\begin{equation}
\ln \mathcal{P}(\mathbf{p} | r_{\rm peak}) = \ln \mathcal{P}(\mathbf{p}) + \ln \mathcal{L}(\mathbf{p}).
\end{equation}

To explore the posterior distribution, we employ MCMC sampling using the \texttt{emcee} package \citep{Foreman2013}, utilizing 500 walkers with 2000 steps each, discarding the first 500 steps as burn-in. The best-fit ellipse parameters are adopted as the median values of the marginalized posterior distributions, with the 16th and 84th percentiles defining the $1\sigma$ confidence intervals. The resulting fit yields $x_c = -0.02^{+0.01}_{-0.01}$, $y_c =0.03^{+0.01}_{-0.01}$, $a =0.50^{+0.02}_{-0.02}$, and $b =0.45^{+0.01}_{-0.01}$. The posterior distributions are presented in Figure \ref{fig:exlup_cornerplot}.  The offset $O_{\mathrm{c}}$ of the ellipse center from the coordinate origin is finally calculated as $O_{\mathrm{c}} = \sqrt{x_c^2 + y_c^2} = 5.6 \pm 1.6$~au.

\bibliographystyle{aasjournal}
\bibliography{references,additional_reference}

\end{document}